\documentclass[tighten, notrackchanges, twocolumn]{aastex631}

\usepackage{amsmath}
\usepackage{amssymb}
\usepackage{amssymb}
\usepackage{mathtools}
\usepackage{mathrsfs}

\usepackage{morefloats}
\usepackage{longtable}
\usepackage{afterpage}

\usepackage{float}

\hypersetup{linkcolor=red,citecolor=blue,urlcolor=magenta}

\usepackage{soul}

\usepackage[makeroom]{cancel}

\graphicspath{{figs/}}

\usepackage{xspace}

\newcommand*{\XPSI}{X-PSI\xspace}
\newcommand*{\NICER}{NICER\xspace}
\newcommand*{\xmm}{XMM-Newton\xspace}
\newcommand*{\MultiNest}{\textsc{MultiNest}\xspace}

\newcommand{\msol}{$M_\odot$\xspace}
\newcommand{\jdbl}{PSR~J0030$+$0451\xspace}
\newcommand{\joh}{PSR~J0740$+$6620\xspace}

\newcommand*\diff{\mathop{}\!\mathrm{d}}

\newcommand{\be}{\begin{equation}}
\newcommand{\ee}{\end{equation}}

\received{2023 March 23}
\revised{2023 July 14}
\accepted{2023 July 15}
\published{2023 October 17}

\submitjournal{The Astrophysical Journal} % Letters}

\shorttitle{Atmospheric effects with NICER}
\shortauthors{Salmi~et~al.}

\begin{document}

\title{Atmospheric Effects on Neutron Star Parameter Constraints with NICER}

\correspondingauthor{T.~Salmi}
\email{t.h.j.salmi@uva.nl}
  
\author[0000-0001-6356-125X ]{Tuomo~Salmi}
\affil{Anton Pannekoek Institute for Astronomy, University of Amsterdam, Science Park 904, 1098XH Amsterdam, the Netherlands}

\author[0000-0003-3068-6974]{Serena~Vinciguerra}
\affil{Anton Pannekoek Institute for Astronomy, University of Amsterdam, Science Park 904, 1098XH  Amsterdam, the Netherlands}

\author[0000-0002-2651-5286]{Devarshi~Choudhury}
\affil{Anton Pannekoek Institute for Astronomy, University of Amsterdam, Science Park 904, 1098XH  Amsterdam, the Netherlands}

\author[0000-0002-1009-2354]{Anna~L.~Watts}
\affil{Anton Pannekoek Institute for Astronomy, University of Amsterdam, Science Park 904, 1098XH  Amsterdam, the Netherlands} 

\author[0000-0002-6089-6836]{Wynn~C.~G.~Ho}
\affil{Department of Physics and Astronomy, Haverford College, 370 Lancaster Avenue, Haverford, PA 19041, USA}

\author[0000-0002-6449-106X]{Sebastien~Guillot}
\affil{Institut de Recherche en Astrophysique et Plan\'{e}tologie, UPS-OMP, CNRS, CNES, 9 avenue du Colonel Roche, BP 44346, F-31028 Toulouse Cedex 4, France}

\author[0000-0002-0428-8430]{Yves~Kini}
\affil{Anton Pannekoek Institute for Astronomy, University of Amsterdam, Science Park 904, 1098XH Amsterdam, the Netherlands}

\author[0000-0002-9407-0733]{Bas~Dorsman}
\affil{Anton Pannekoek Institute for Astronomy, University of Amsterdam, Science Park 904, 1098XH Amsterdam, the Netherlands}

\author[0000-0003-4357-0575]{Sharon~M.~Morsink}
\affil{Department of Physics, University of Alberta, 4-183 CCIS, Edmonton, AB, T6G 2E1, Canada}

\author[0000-0002-9870-2742]{Slavko~Bogdanov} 
\affil{Columbia Astrophysics Laboratory, Columbia University, 550 West 120th Street, New York, NY 10027, USA}

\begin{abstract}
We present an analysis of the effects of uncertainties in the atmosphere models on the radius, mass, and other neutron star parameter constraints for the NICER observations of rotation-powered millisecond pulsars. To date, NICER has applied the X-ray pulse profile modeling technique to two millisecond-period pulsars: PSR J0030+0451 and the high-mass pulsar PSR J0740+6620. 
These studies have commonly assumed a deep-heated, fully ionized hydrogen atmosphere model, although they have explored the effects of partial-ionization and helium composition in some cases. 
Here, we extend that exploration and also include new models with partially ionized carbon composition, externally heated hydrogen, and an empirical atmospheric beaming parameterization to explore deviations in the expected anisotropy of the emitted radiation. 
None of the studied atmosphere cases have any significant influence on the inferred radius of PSR J0740+6620, possibly due to its X-ray faintness, tighter external constraints, and/or viewing geometry. 
In the case of PSR J0030+0451, both the composition and ionization state could significantly alter the inferred radius. 
However, based on the evidence (prior predictive probability of the data), partially ionized hydrogen and carbon atmospheres are disfavored.
The difference in the evidence for ionized hydrogen and helium atmospheres is too small to be decisive for most cases, but the inferred radius for helium models trends to larger sizes around or above 14--15 km.
External heating or deviations in the beaming that are less than $5\,\%$ at emission angles smaller than $60\degr$, on the other hand, have no significant effect on the inferred radius.
\end{abstract}

%\keywords{radiative transfer -- methods: numerical -- pulsars: general -- stars: atmosphere -- stars: neutron -- X-rays: stars}
 
\section{Introduction}
\label{sec:intro}

Mass and radius measurements of neutron stars (NSs) can be used to probe the strongly degenerate matter at supranuclear densities in NS cores. 
This is because the NS mass-radius relation depends sensitively on the pressure-density relation, i.e., the equation of state of the interior matter \citep[see, e.g.,][]{LP01,Ozel2009,Lattimer12ARNPS,Hebeler2013,Baym2018}.
One useful approach to measure the masses and radii of different NSs is by modeling the observed X-ray pulses produced by \textit{hot spots} on the surfaces of rotating NSs. 
This extensively studied technique exploits the general and special relativistic effects modifying the pulse shape depending on mass, radius, and other model parameters \citep[see, e.g.,][]{PFC83,Strohmayer1992,ML98,PG03,MLC07,lomiller13,AGM14,ML15,Watts2016,NP18,BLM_nicer19}. 

In recent years, the pulse profile modeling method has been applied by NASA’s Neutron Star Interior Composition Explorer \citep[\NICER;][]{Gendreau2016}, concentrating on rotation-powered millisecond pulsars (RMPs). 
In these pulsars, the hot regions on the NS surface are heated by the flow-back of electrons or positrons that are accelerated in the magnetosphere (in contrast to, e.g., hot spots in accreting millisecond pulsar, which are heated due to the accretion from a companion star). 
The bombarding particles are produced in single photon magnetic pair cascades originating at the open field line region \citep[see, e.g.,][]{RudermanSutherland1975,arons81,HM01}. 
So far, results for two pulsars have been reported: \jdbl \citep{MLD_nicer19,RWB_nicer19,vinciguerra2023bravo} and \joh \citep[][]{Miller2021,Riley2021,salmi2022}. 
These studies have started to restrict dense matter models \citep[see, e.g.,][]{Raaijmakers2021} and pointed to nonantipodal magnetic polar cap geometries \citep[see, e.g.,][]{BWH_nicer19}.

In the studies mentioned above, the energy and emission angle dependence (i.e., beaming pattern) of the radiation escaping from the NS surface is calculated from the input physics (for a given effective temperature\footnote{Effective temperature $T_{\mathrm{eff}}$ is the temperature of a blackbody producing a bolometric photon flux equal to the bolometric photon flux from the atmosphere model.} and surface gravity) based mainly on the \texttt{NSX}  deep-heated, nonmagnetic, and fully ionized hydrogen atmosphere model \citep{Ho01}. 
Sensitivity to chemical composition was tested using the \texttt{NSX} fully ionized helium atmosphere for \jdbl in \citet{MLD_nicer19} and for \joh in \citet{Riley2021}. 
In addition, the sensitivity to \texttt{NSX} with partially ionized hydrogen was checked for \joh in \citet{Miller2021}. 
The alternative atmosphere cases resulted in consistent mass and radius values, except in the case of helium for \jdbl, which was discarded because of the very high inferred NS mass around $2.7$ \msol, with most of the posterior probability above $2.0$ \msol \citep{MLD_nicer19}. 
In this work, we explore additional cases with a helium atmosphere and show that for \jdbl well-fitting solutions with helium atmospheres can be found also with more realistic NS masses (see discussion in Section \ref{sec:disc_composition}). 

The uppermost atmospheric layers, which determine the properties of escaping radiation, are usually expected to consist of hydrogen due to the rapid sinking of heavier elements via diffusive gravitational separation \citep{AI1980,Hameury1983,Brown2002,ZP2002} on a timescale of $\sim 1\textrm{--}100$ s \citep{Romani1987}. 
However, they could consist of helium if all the hydrogen originally present has been converted to helium via diffusive nuclear burning \citep{Chang&Bildsten2004,Wijngaarden2019}, or if the accretion that recycled the RMP happened from a completely hydrogen-depleted companion star \citep[see][and references therein]{Bogdanov2021}. 
Less likely, but still possible, would be an atmosphere consisting of even heavier elements, e.g., due to nuclear burning, a wind exposing the underlying heavier elements, or a lack of light element accretion \citep{Chang&Bildsten2004,Bogdanov2021}. 
We therefore also perform some new analysis using a partially ionized carbon \texttt{NSX} atmosphere model. 

So far, the inferred effective temperatures in RMPs analyzed by \NICER have been around $\log_{10} (T_{\mathrm{eff}}/K) \sim 6$ for which more than 99\,\% of the hydrogen is completely ionized at all depths \citep[see Figure 4 of][]{ZPS96}. 
However, neutral atoms could still exist in dense enough layers of the atmosphere, or there could be temperature layers cold enough to produce small bound--bound and bound--free opacity features in the escaping radiation \citep{ZPS96}.   
Therefore, we have also explored more partially ionized models. 
The caveat, however, is that the opacity tables needed for partially ionized models do not
cover the entire range of energies, temperatures, and densities that are relevant in RMPs \citep{IglesiasRogers1996,Badnell2005,Colgan2016,Bogdanov2021}. 

Besides the composition and the ionization state, a few other possible sources of uncertainties in the atmosphere models were also discussed in \citet[][for a review see also, e.g., \citealt{Potekhin2014}]{Bogdanov2021}. 
These include the magnetic field strength, depth of energy deposition of the bombarding particles, and the accuracy of the computational methods used in atmosphere modeling. 
These effects were qualitatively predicted to produce deviations that are small for the \NICER sources analyzed so far, but we quantify this for the cases of deposition depth and computational methods by performing pulse profile analyses using an independent numerical implementation of the atmosphere model, assuming an externally heated, fully ionized hydrogen atmosphere from \citet[][see also \citealt{BPO19}]{Salmi20}. 
In \citet{Bogdanov2021} comparisons were previously done against another atmosphere code called McPHAC \citep{Haakonsen2012}, identifying a couple of issues in using the latter. 
Finally, in order to allow small deviations from the nominal beaming in any of the atmosphere models, we also study new pulse profile models with additional free parameters that can modify the beaming. 

The remainder of this paper is structured as follows.
In Section \ref{sec:model}, we present the methods that we use to estimate the atmospheric uncertainties in pulse profile modeling and parameter constraints. 
We introduce the different model atmospheres and beaming correction formalism that are applied in our analysis.   
In addition, we compare the spectra and beaming patterns calculated using different atmosphere models. 
In Section \ref{sec:results}, we present the parameter constraints obtained with different atmosphere model assumptions, considering first \joh  and then \jdbl. 
We discuss the implications of our results for future analysis in Section \ref{sec:discussion} and conclude in Section \ref{sec:conclusions}.
 
\section{Modeling procedure}\label{sec:model}

In this Section, we present the methods we use to produce the X-ray pulses and fit the data. 
We focus first on the new features applied for the atmosphere modeling, and then summarize the other modeling aspects, which are mainly shared with those from \citet{BLM_nicer19,RWB_nicer19,Riley2021,salmi2022,vinciguerra2023sim,vinciguerra2023bravo}. 
For all posterior computations, we use the X-ray Pulse Simulation and Inference\footnote{\url{https://github.com/xpsi-group/xpsi}} (\XPSI) code, with versions ranging from \texttt{v0.7.3} to \texttt{v1.2.1} \citep{xpsi}.\footnote{The versions are practically identical for the considered models; the only actual difference is the fix of a numerical ray tracing issue since v0.7.12, affecting only a few parameter vectors.} 
The exact version for each run, data products, posterior sample files, a complete set of output figures, and all of the analysis files in the Python language may be found in the persistent Zenodo repository of \citet{salmi_zenodo23}. 
We use a new atmosphere extension that was developed for the free beaming runs and introduced in v1.2.0 of \XPSI;  in \citet{salmi_zenodo23}, we also provide instructions on how to install that for the previous versions of \XPSI to allow reproducibility of this work.    

\subsection{Atmosphere Modeling}

We employ atmosphere models to determine the specific intensity $I(E,\mu)$ emitted from the NS surface, where $E$ is the photon energy and $\mu$ is the cosine of the emission angle with respect to the surface normal. 
Since calculating this quantity is computationally expensive, we use pre-computed look-up tables of $I(E,\mu)$ when calculating the pulse profiles, similar to the previous \NICER studies. The tables used in this study are presented next
in Section \ref{sec:num_atmos}, and the procedure for allowing deviation from the precalculated beaming pattern is described in Section \ref{sec:beaming_formalism}.

\subsubsection{Numerical Model Atmospheres}\label{sec:num_atmos}
 
We use 5 different look-up tables for the numerical atmosphere models. 
The first is the deep-heated and fully ionized hydrogen \texttt{NSX} model \citep[nsxH;][]{Ho01}, used in the previous \XPSI analyses. 
This table provides the logarithm of the ratio of the specific intensity with respect to the cube of the effective temperature, $(I/T_{\rm eff}^3)$, for a grid in the logarithm of the effective temperature, $\log_{10} (T_{\rm eff}/\mathrm{K})$, the logarithm of the surface gravity, $\log_{10} (g_{\mathrm{s}}/\mathrm{cm\,s}^{-2})$, the cosine of the emission angle, $\mu$, and the logarithm of photon energy with respect to the temperature, $\log_{10}(E/kT_{\rm eff})$ \citep[see][for more details]{Bogdanov2021}. 
The ranges and the number of grid points for each atmosphere model parameter are shown in Table \ref{table:num_atmos}.
The same information is also displayed for the other atmosphere tables used in this paper. 

\begin{deluxetable*}{lcccc}
\setlength{\tabcolsep}{19pt}
\tablewidth{0pt}
\tablecaption{Summary of the Numerical Atmosphere Tables}
\tablehead{\colhead{Model} & \colhead{$\log_{10} (T_{\rm eff}/\mathrm{K})$} & \colhead{$\log_{10} (g_{\mathrm{s}}/\mathrm{cm\,s}^{-2})$} & \colhead{$\log_{10}(E/kT_{\rm eff})$} & \colhead{$\mu$} }
\startdata
hatmH & $[5.1,6.8]: 18$  & $[13.7,14.6]: 10$ & $[-1.3,2.0]: 166^{\mathrm{a}}$ & $[0.0, 1.0]$: 11 \\
\hline
nsxCp & $[5.6,6.7]: 23$ & $[13.6,15.0]: 15$ & $[-1.3,2.0]: 166$ & $[1\times10^{-6}, 1 -1 \times 10^{-6}]$: 67 \\
\hline
nsxHp & $[5.2,6.7]: 31$ & $[13.6,15.0]: 15$ & $[-1.3,2.0]: 166$ & $[1\times10^{-6}, 1 -1 \times 10^{-6}]$: 67 \\
\hline
nsxHe & $[5.1,6.5]: 29$ & $[13.7,14.7]: 11$ & $[-1.3,2.0]: 166$ & $[1\times10^{-6}, 1 -1 \times 10^{-6}]$: 67 \\
\hline
nsxH & $[5.1,6.8]: 35$& $[13.7,15.0]: 14$ & $[-1.3,2.0]: 166$ & $[1\times10^{-6}, 1 -1 \times 10^{-6}]$: 67 \\
\enddata
\tablecomments{\ \ The atmosphere grid ranges ([$\cdots,\cdots$]) and number of grid points (:$\cdots$) are shown for fully ionized externally heated hydrogen (hatmH), partially ionized carbon (nsxCp), partially ionized hydrogen (nsxHp), fully ionized helium (nsxHe), and fully ionized hydrogen (nsxH).
For \texttt{NSX} models, emission angles were placed at every $1.\degr5$, with three additional values near $\mu=0$, and $\mu=1$. In case of hatmH, the grid points were based on the sample points of Gauss-Legendre quadrature between $0$ and $1$. 
\\
$^{\mathrm{a}}$ Originally a logarithmically spaced grid of $E$ from $4.1 \times 10^{-4}$ to $41$ keV with $360$ grid points was used for calculating this model, but the intensity values were then interpolated to the same $\log_{10}(E/kT_{\rm eff})$ grid as in \texttt{NSX}, for simplicity. \\
}
\end{deluxetable*}\label{table:num_atmos}

The second look-up table is the deep-heated and fully ionized helium NSX model \citep[nsxHe;][]{Ho01}, included also, e.g., in the analysis of \citet{Riley2021}.
This table is otherwise similar to the hydrogen table, except for having slightly smaller upper limits for high temperature and surface gravity. 

For the third and fourth look-up tables, we use the deep-heated partially ionized hydrogen \texttt{NSX} model (nsxHp), applied previously in \citet{Miller2021},\footnote{Note that the \NICER analyses of \citet{MLD_nicer19,Miller2021} do not use \XPSI but rather an independent ray-tracing and inference code.}  and the partially ionized carbon \texttt{NSX} model, \citep[nsxCp, which is essentially the same as in][except for different grids used in the calculation]{HH09}.
These tables have roughly similar ranges as the previous ones, as seen in Table \ref{table:num_atmos}, although the carbon model does not reach such cold temperatures as the others. 

For the fifth look-up table, we use an externally heated
and fully ionized hydrogen model (hatmH) from \citet{Salmi20}.
This model is mostly similar to \texttt{NSX}  but it allows the bombarding return-current particles to release their kinetic energy as heat at different depths of the atmosphere instead of assuming that all the heat is released at very deep layers. 
For RMPs, deep heating is expected to be a reasonable assumption, as explained in Section \ref{sec:disc_other}, but we nevertheless check if relaxing this assumption (adopting an extreme case) might affect the results. 
In addition to external heating, the new model calculates the photon--electron scattering using the exact Compton scattering formalism and a temperature iteration scheme similar to those in \citet{SPW12,SSP19,Suleimanov20}, which differ slightly from the Thomson scattering and iteration schemes used in \texttt{NSX}. 

In this paper, the hatmH atmosphere was assumed to be heated by particles with a power-law energy distribution, characterized by a minimum Lorentz factor $\gamma_{\mathrm{min}}=10$, power-law slope $\delta = 2$, and a maximum Lorentz factor $\gamma_{\mathrm{max}}= 2 \times 10^{5}$. 
These choices correspond to the fiducial parameter setup in \citet{Salmi20}, and they lead to an average Lorentz factor $\gamma_{\mathrm{avg}} = 100$ of a bombarding particle. 
These values were selected to maximize the difference with the deep-heated atmosphere model, whilst not being incompatible with our understanding of pulsar physics, given the uncertainties in simulations (see discussion in Section \ref{sec:disc_other}). 
Due to the computationally more expensive atmosphere model, fewer grid points were produced for the hatmH table in effective temperature, surface gravity, and emission angle, as seen in Table \ref{table:num_atmos}. 

\subsubsection{Beaming Correction Formalism and Priors}\label{sec:beaming_formalism}

In some of our models (see Tables \ref{table:run_summary_2} and \ref{table:run_summary} for details), we introduce additional freedom in the beaming pattern predicted by the numerical atmosphere models. 
This is done by multiplying the intensity values $I(E,\mu)_{\mathrm{NUM}}$ (interpolated from the atmosphere table) with a polynomial function modifying the beaming:
\be\label{eq:Iemu_fbemu}
I(E,\mu,a,b,c,d) = I(E,\mu)_{\mathrm{NUM}}f(E,\mu,a,b,c,d),
\ee
where
\be\label{eq:fbcene}
f(E,\mu,a,b,c,d) = C\left( 1+a\left( \frac{E}{\mathrm{keV}}\right)^{c}\mu+b\left( \frac{E}{\mathrm{keV}}\right)^{d} \mu^{2}\right),
\ee
$a$, $b$, $c$, and $d$ are beaming parameters.
In the case of an energy-independent beaming correction, we fix $c=0$ and $d=0$, but in the case of an energy-dependent correction, we keep all of them free. 
The constant $C$ is determined so that $\int_{0}^{1}\mu f(\mu,a,b,c,d) \diff \mu  = 1/2$, which gives:
\be\label{eq:cnorm2}
C = \frac{1}{1 + (2/3)a(E/\mathrm{keV})^{c} + (1/2)b(E/\mathrm{keV})^{d}}.
\ee
This constant is chosen to minimize the change in the total flux spectrum $F(E) = \int_{0}^{1}\mu I(E, \mu,a,b,c,d) \diff \mu$ for any given $a$, $b$, $c$, and $d$. 
This choice precisely conserves the emergent spectrum when the model intensity $I(E,\mu)_{\mathrm{NUM}}$ does not depend on the emission angle. 
We found that this normalization factor already eliminates most of the difference in the spectrum caused by non-zero values of  $a$, $b$, $c$, and $d$, leaving the spectrum unchanged with better than one percent accuracy for the beaming corrections we allow in our prior support.
In addition, we performed a couple of low-resolution test runs setting either $C=1$ or finding the value of $C$ by numerical integration, requiring that $\int_{0}^{1} \mu I(E,\mu)_{\mathrm{NUM}}f(\mu,a,b,c,d) \diff\mu = \int_{0}^{1} \mu I(E,\mu)_{\mathrm{NUM}}\diff\mu$ separately for each emitted photon (increasing the computational cost significantly). 
We found that the inferred radii do not vary by more than the 1 km fluctuation caused by the low-resolution sampling settings (see Section \ref{sec:posterior_computation}). 
This is probably a result of the inferred beaming parameters being close to zero, leading $C$ to be close to unity.

In the inference runs performed in this work, the prior support for the beaming parameters was based on a condition that the relative correction in the intensity (i.e., $|(f(E,\mu)-1)| $) needs to remain below a pre-defined threshold (otherwise the sample is rejected).
For \joh we required the correction to be at most 10\,\% at any energy and emission angle. 
For \jdbl we required the correction to be at most 5\,\% at any energy and emission angle below $60\degr$ (setting no limits for angles higher than that).
\footnote{These checks were done for 100 (or 50) emission angles linearly spaced in $\mu = [0.0, 1.0]$ (or $\mu = [0.5, 1.0]$) and for 100 energies linearly spaced in  $\log_{10}(E/kT_{\rm eff}) = [-1.5, 2.2]$, using $T_{\rm eff}$ of both the primary and secondary hot regions.}\textsuperscript{,}
\footnote{
These conditions produce an anticorrelation between $a$ and $b$, since a constant \textit{noncorrecting} beaming function is found more easily if these factors have opposite signs so that the $\mu$ and $\mu^{2}$ terms cancel each other. 
For more efficient sampling, these priors can be mimicked by drawing $a$ and $b$ from normal distributions centered around $0$ and $-a$, respectively.
However, for simplicity, only initially uniform distributions were used for the final runs reported here.
}

In both cases, the maximum allowed mean correction (averaged over all energies and angles) was around $6-9$\,\%.
These limits were chosen to avoid too much freedom in the parameter space, and they were roughly based on the differences between atmosphere models shown in Section \ref{sec:model_comparison}.

In the case of \joh, the bounds of the beaming parameters $a$ and $b$ were set to $[-1.0,1.0]$ while keeping $c$ and $d$ fixed to zero, and in case of \jdbl the bounds for all $a$, $b$, $c$, and $d$ were set to $[-0.7,0.7]$.
These ranges were found to always include at least $85\,\%$ of the prior space of each parameter.

For simplicity, we also assumed the beaming parameters to be shared between all the emitting regions, even though the regions could have different temperatures and the beaming correction might depend on the temperature. 

\subsection{Pulse Profile Modeling Using X-PSI}\label{sec:xpsi_modeling}
 
We calculate the energy-resolved X-ray pulses using the same pulse profile modeling technique, relying on the `Oblate Schwarzschild' approximation, as was done in the previous \NICER analyses  \citep[see, e.g.,][and the references in Section \ref{sec:intro}]{BLM_nicer19}. 
Our choices regarding the model parameters (besides the beaming parameters explained in Section \ref{sec:beaming_formalism}), prior distributions, and likelihood calculation, follow those used in \citet{vinciguerra2023sim, vinciguerra2023bravo} for \jdbl, and those used in \citet{salmi2022} for \joh.
For example, in both cases, we allow multiple imaging for the relativistic ray-tracing (as explained in \citealt{Riley2021}). 
A few more features of the models are described next. 
 
As in the previous works, for \joh we use the \texttt{ST-U} (\textit{Single-Temperature-Unshared}) hot region model, with two circular regions having uniform (but different) effective temperatures of the atmosphere. 
For \jdbl, the headline results from \citet{RWB_nicer19} were produced using
\texttt{ST+PST} (\textit{Single-Temperature+Protruding-Single-Temperature}), in which one hot region is a circle, and the other is allowed to form ring and crescent-like structures (both with uniform
but different temperatures). 
However, the \texttt{ST+PST} model is computationally expensive, so it is not feasible to fully explore the effects of atmosphere choices using \texttt{ST+PST}.
Instead, we first performed an exhaustive exploration of the effects of different atmosphere models on \jdbl using the much simpler \texttt{ST-U} spot model, which is the simplest tested model that does not show any structures in the residuals. 
This exploration allows us to find the cases where the \texttt{ST-U} results for \jdbl are most sensitive to atmosphere choices. 
These most sensitive cases were selected for further analysis with the more complex \texttt{ST+PST} spot model.

When using \texttt{ST+PST}, we reduce the resolution in hot region cells, phase bins in the star frame, and energy bins for specific photon flux, as in the low-resolution runs in \citet[][see Section 2.3.1 there for details of the settings]{vinciguerra2023sim}. 
This is done to mitigate the extra computational expense caused by the more complex model, while still allowing the use of high enough sampling resolution of the parameter space (see Section \ref{sec:posterior_computation}). 

For the instrument response model, we use only energy-independent effective area scaling factors for both \joh and \jdbl. We combined the uncertainty in the distance and the scaling factor into a parameter called $\beta$ for the latter \citep[as in][]{vinciguerra2023sim}, except when jointly fitting \NICER and \xmm data. 
Energy-dependent scaling factors were originally used in \citet{RWB_nicer19} but were later replaced with energy-independent ones for the newer \jdbl and \joh analysis as explained in \citet{vinciguerra2023sim}. 

The calculation of the background-marginalized likelihood function is similar to that used in the previous works.  For \jdbl, we apply background constraints using \xmm \citep[similarly to][]{vinciguerra2023bravo} for atmosphere cases that showed large differences in the radius when \xmm was not included.  For \jdbl, the background is typically inferred to include most of the unpulsed emission \citep{MLD_nicer19,RWB_nicer19} if one does not use any constraints; with \xmm, we can limit the \NICER background to be smaller (by constraining the number of source photons). 
For all of our \joh analyses, we apply the nonsmoothed \NICER space weather background estimate as a lower limit for the background without using the \xmm data \citep[corresponding to the case labeled W21-0.9xSW in][]{salmi2022}. This was shown to give similar results as constraining the background using \xmm for \joh in \citet{salmi2022}.

\subsection{Data Sets}

In this section, we summarize the event data sets used in the analysis of this paper.

\subsubsection{\joh}
 
For \joh analysis, we used the same pre-processed \NICER X-ray event data \citep[described in][]{Wolff21} and instrument response files as in \citet{Miller2021}, \citet{Riley2021}, and we partly used the same data as in \citet{salmi2022}. 
Instead of including \xmm data in the analysis, we used a lower limit for the \NICER background based on the space weather background estimator \citep{SpaceWeather}, as in some models of \citet{salmi2022}.
 
\subsubsection{\jdbl} 

For \jdbl analysis, we used the same pre-processed X-ray event data and instrument response files as in \citet{vinciguerra2023bravo}. 
The \NICER data set is thus similar to that introduced in \citet{BGR_nicer19}, and used in \citet{RWB_nicer19}, \citet{MLD_nicer19}, but updated adopting the most recent \NICER instrument response, as in \citet{Riley2021}, \citet{salmi2022}. 
In addition, for some analyses, we also used the same \xmm data as introduced in \citet{BG2009}, \citet{vinciguerra2023bravo}.  

\subsection{Posterior Computation}\label{sec:posterior_computation}
 
As in the previous works, we compute the posterior samples using \MultiNest \citep{MultiNest_2008,multinest09,PyMultiNest}. 
For details of our nested sampling protocol, we refer to \citet[][especially the appendices]{RWB_nicer19}, \citet[][especially chapter 3 and the associated appendix]{riley_thesis}.  
The resolution settings for sampling were chosen to be adequate for an exploratory analysis, where the main focus is to compare the effects of different atmosphere model choices. 

For the \joh analysis, we employed the `low-resolution' settings from \citet{Riley2021}: $4\times10^{3}$ live points; and $0.1$ sampling efficiency\footnote{
We note that this factor is not the number set as the \MultiNest sampling efficiency parameter, since it is first corrected by \XPSI to account for the non-uniform prior hypercube volume as explained in \citet{riley_thesis}.}; and an estimated remaining log-evidence of $10^{-1}$. 
This was found to produce stable results that do not significantly change from run to run with different random seeds for sampling with \MultiNest.
However, as noted in \citet{Riley2021}, \citet{salmi2022}, using 10 times more live points would typically slightly broaden the credible intervals and shift the median radius up by 0.2--0.4 km.

For the \jdbl analyses, we used slightly different resolution settings; $1\times10^{4}$ live points; and $0.3$ sampling efficiency. 
These settings were chosen to make the computation reasonably fast (noting that \jdbl analysis is more expensive than \joh analysis with the same settings, likely due to the higher number of detected counts), while avoiding fluctuation in the results for runs with different \MultiNest seeds. 
The latter was found to be a problem especially for our initial runs with additional beaming parameters, where $4\times10^{3}$ live points and $0.8$ sampling efficiency (which seemed sufficient without the beaming parameters) resulted in the inferred NS radius varying by around 1 km between identical runs and always being biased toward higher radii compared to the run with higher sampling resolution. 
Therefore, the higher resolution was used in all the results reported in this paper. 
In most of our runs, the mode-separation variant of \MultiNest (known also as the multimodal option) was deactivated (i.e., isolated modes were not evolved independently), except for the joint \NICER and \xmm \texttt{ST-U} and \NICER-only \texttt{ST+PST} fits of \jdbl. 
The settings used for each run are listed in Tables \ref{table:run_summary_2} and \ref{table:run_summary}.

\subsection{Model Comparison}\label{sec:model_comparison}

We start by comparing emergent spectra and beaming patterns predicted by the different atmosphere models used in this paper. 
For the comparison, we select the effective temperature (of the hottest region) and surface gravity values that correspond to the maximum likelihood parameter vector inferred when applying the fully ionized hydrogen \texttt{NSX} \texttt{ST-U} model (model nsxH from Table \ref{table:num_atmos}) for either \jdbl or \joh; these are $\log_{10} (T_{\rm eff}/\mathrm{K}) = 6.1059$, $\log_{10} (g_{\mathrm{s}}/\mathrm{cm\,s}^{-2}) = 14.143$ and $\log_{10} (T_{\rm eff}/\mathrm{K}) = 6.0409$, $\log_{10} (g_{\mathrm{s}}/\mathrm{cm\,s}^{-2}) = 14.194$ respectively. 
In addition, we also show one higher temperature example with $\log_{10} (T_{\rm eff}/\mathrm{K}) = 6.4$ and $\log_{10} (g_{\mathrm{s}}/\mathrm{cm\,s}^{-2}) = 14.15$. 
The different spectra are presented in Figure \ref{fig:spectrum} for $\mu=0.5$, i.e., for a $60 \degr$ emission angle. 
This figure shows only small deviations between the different models when considering the relatively low temperatures typical for \jdbl or \joh (the difference in temperatures between the two is small). 
Only the partially ionized carbon model predicts a significantly different spectrum with several lines and edges \citep[as in, e.g.,][]{HH09,SKPW14}. 
The impact of external heating starts to become major (larger than that between hydrogen and helium composition) only at higher temperatures, such as in the $\log_{10} (T_{\rm eff}/\mathrm{K}) = 6.4$ case shown in the lower panel of Figure \ref{fig:spectrum}. 
There the partially ionized hydrogen spectrum also starts to deviate more from fully ionized hydrogen. 
Since features in opacity, such as resonances due to bound states, can cause absorption of photons and heat the outer atmospheric layers in partially ionized models, it is not too surprising that both partial ionization and external heating seem to have a similar effect on the spectrum, increasing the low-energy emission.

%\figsetstart
%\figsetnum{1}
%\figsettitle{Spectrum comparison}

%\figsetgrpstart
%\figsetgrpnum{1.1}
%\figsetgrptitle{Spectra for J0030}
%\figsetplot{f1_1.png}
%\figsetgrpnote{The emergent spectra for different atmosphere models at $60 \degr$ emission angle. Temperature and surface gravity correspond to the best-fit results (for the hottest region) from \jdbl analysis with the ST-U-nsxH model.}
%\figsetgrpend

%\figsetgrpstart
%\figsetgrpnum{1.2}
%\figsetgrptitle{Spectra for J0740}
%\figsetplot{f1_2.png}
%\figsetgrpnote{ The emergent spectra for different atmosphere models at $60 \degr$ emission angle. Temperature and surface gravity correspond to the best-fit results (for the hottest region) from \joh analysis with the ST-U-nsxH model.}
%\figsetgrpend

%\figsetgrpstart
%\figsetgrpnum{1.3}
%\figsetgrptitle{Spectra for high temperature (T=6.4)}
%\figsetplot{f1_3.png}
%\figsetgrpnote{ The emergent spectra for different atmosphere models at $60 \degr$ emission angle. Temperature and surface gravity are fixed to $\log_{10} (T_{\rm eff}/\mathrm{K}) = 6.4$ and $\log_{10} (g_{\mathrm{s}}/\mathrm{cm\,s}^{-2}) = 14.15$). }
%\figsetgrpend

%\figsetend

{
    \begin{figure}[t!]
    \centering
    \resizebox{\hsize}{!}{
    \includegraphics[
    width=\textwidth]{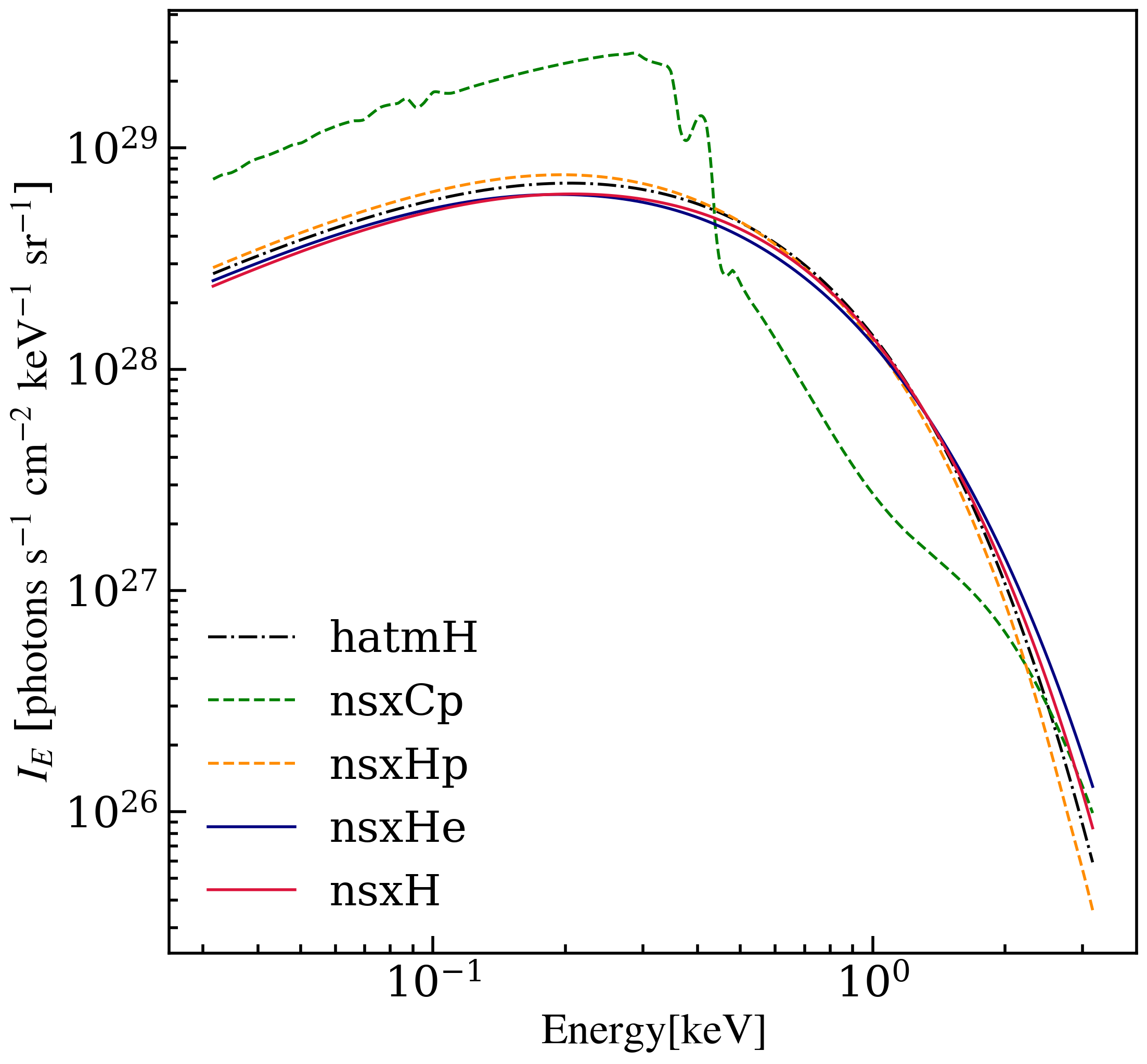}
    }
    \resizebox{\hsize}{!}{
    \includegraphics[
    width=\textwidth]{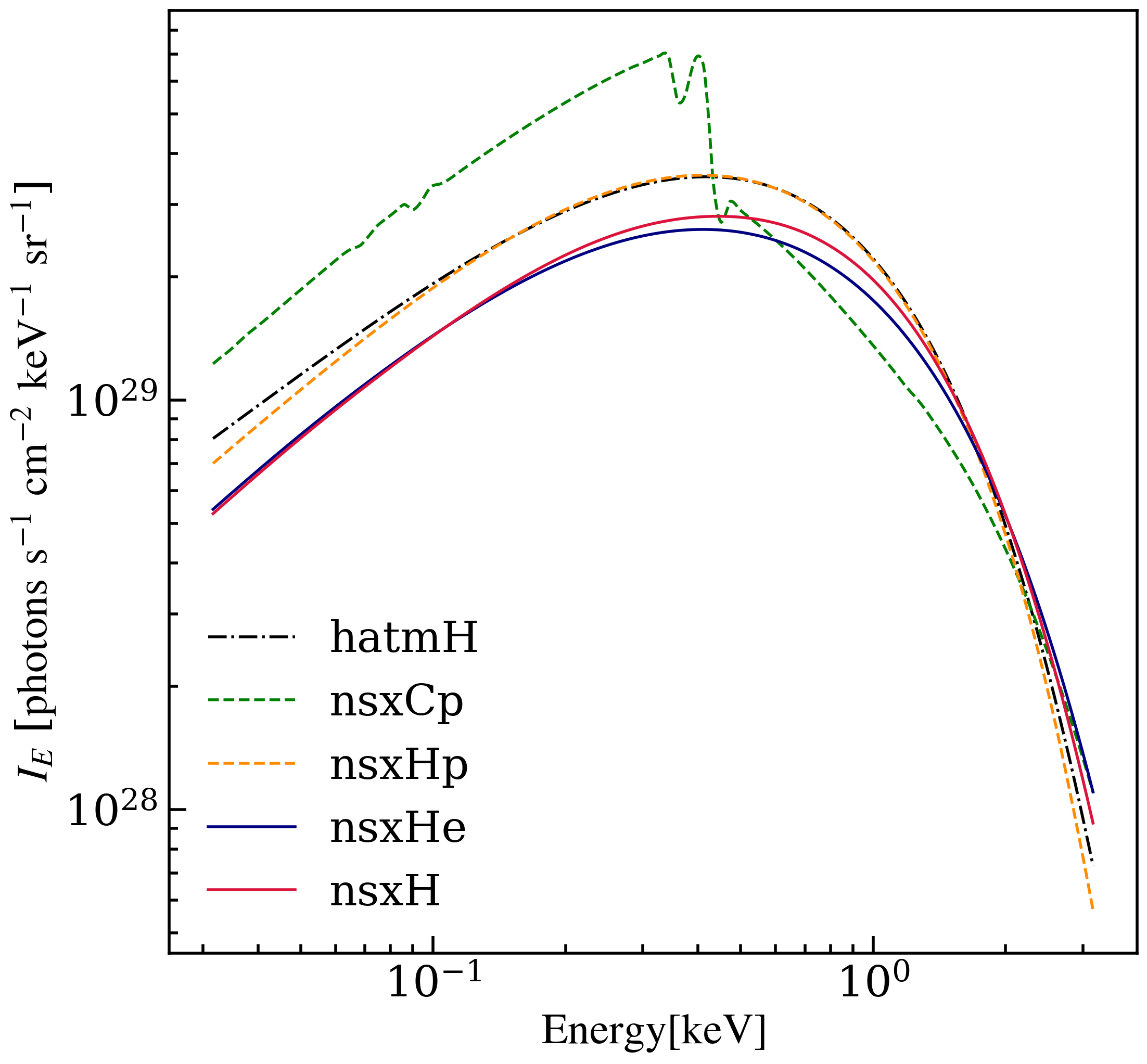}
    }    
    \caption{\small{
    \textit{Upper panel}: the emergent spectra for different atmosphere models at $60 \degr$ emission angle. 
    Temperature and surface gravity correspond to the best-fit results (for the hottest region) from \jdbl analysis with the ST-U-nsxH model ($\log_{10} (T_{\rm eff}/\mathrm{K}) = 6.1059$, $\log_{10} (g_{\mathrm{s}}/\mathrm{cm\,s}^{-2}) = 14.143$). 
    The different atmosphere models are explained in Table \ref{table:num_atmos}.\\
    \textit{Lower panel}: the same as the upper panel, except with a higher temperature $\log_{10} (T_{\rm eff}/\mathrm{K}) = 6.4$, and surface gravity fixed to $\log_{10} (g_{\mathrm{s}}/\mathrm{cm\,s}^{-2}) = 14.15$. 
    The complete figure set (3 images), including also the spectra for best-fit parameters from \joh analysis, is available in the online journal (HTML version). 
    }}
    \label{fig:spectrum}
    \end{figure}
}

{
    \begin{figure}[t!]
    \centering
    \resizebox{\hsize}{!}{\includegraphics[
    width=\textwidth]{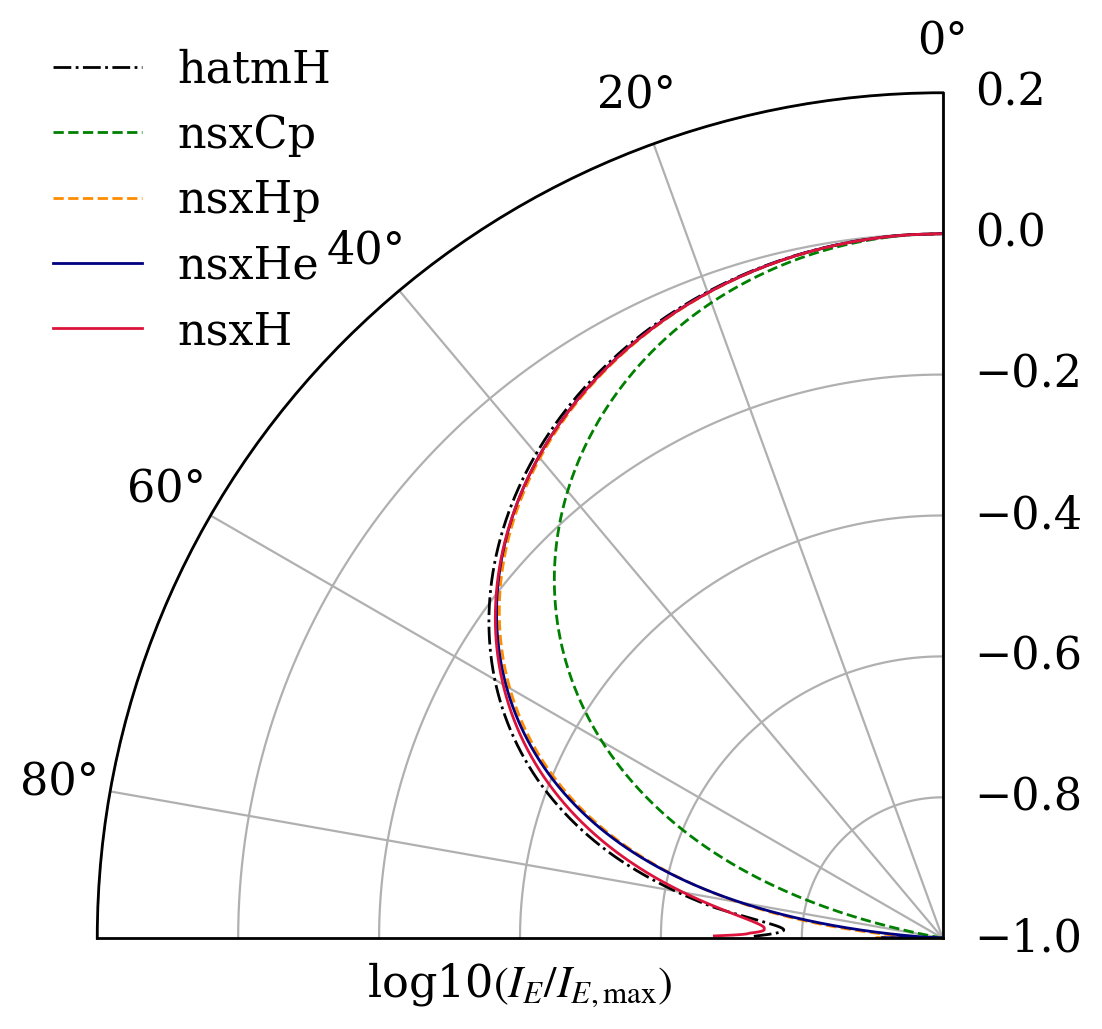}}
    \resizebox{\hsize}{!}{\includegraphics[
    width=\textwidth]{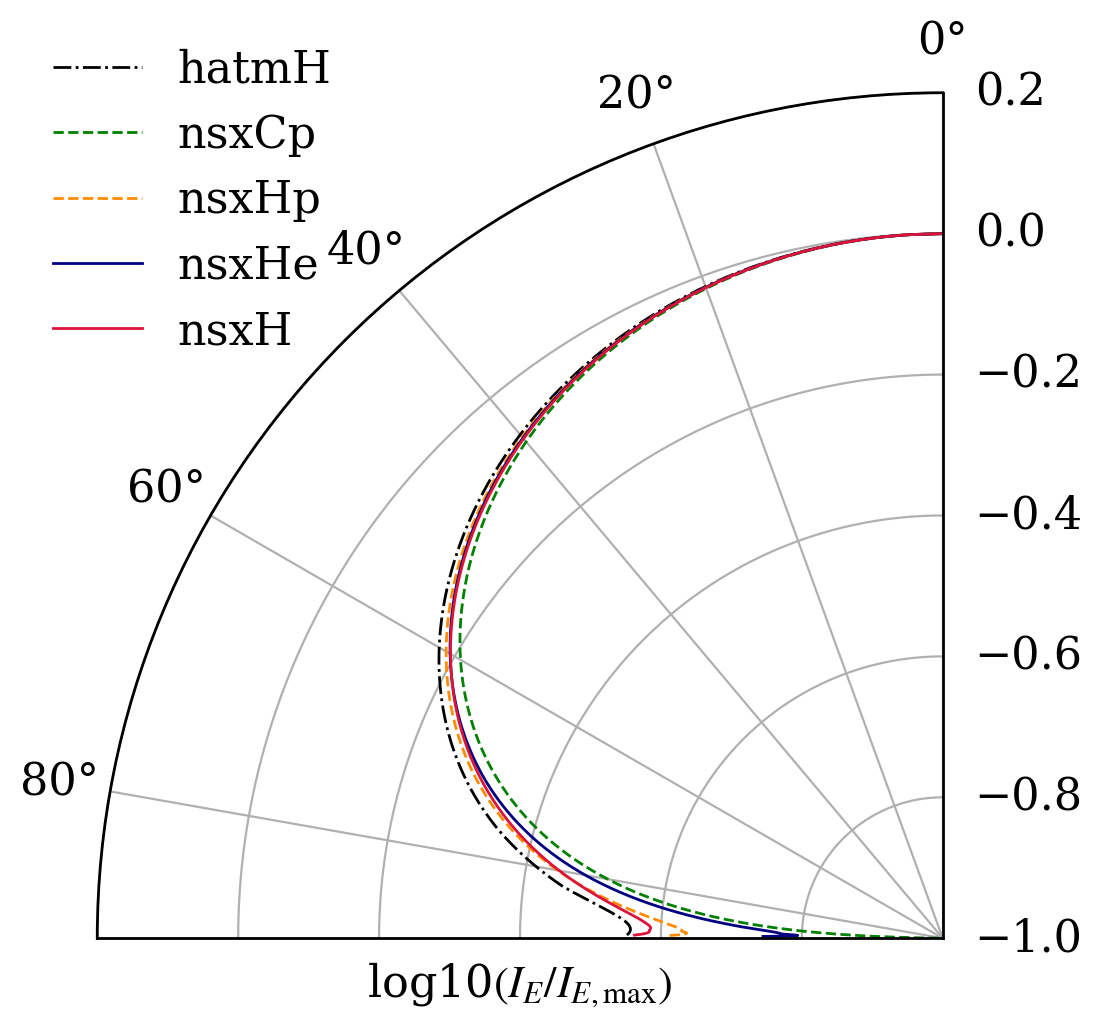}}    
    \caption{\small{
    \textit{Upper panel}: 
    the beaming patterns at 1.0 keV with temperature and surface gravity corresponding to the best-fit results from \jdbl analysis with the ST-U-nsxH model. See Figure \ref{fig:spectrum} and Table \ref{table:num_atmos} for more details. \\ 
    \textit{Lower panel}:
    the same as the upper panel, except with a higher temperature $\log_{10} (T_{\rm eff}/\mathrm{K}) = 6.4$, and surface gravity fixed to $\log_{10} (g_{\mathrm{s}}/\mathrm{cm\,s}^{-2}) = 14.15$. 
    The complete figure set (5 images), including the beaming pattern for 0.5 keV and 2.0 keV with \jdbl parameters, and the pattern at 1.0 keV with \joh parameters is available in the online journal (HTML version).
    }}
    \label{fig:beaming}
    \end{figure}
}

The different beaming patterns are presented in Figure \ref{fig:beaming}. The beaming patterns for all of the atmosphere models except for the carbon model are very similar. 
We also see that the difference caused by composition or partial ionization is larger than the difference caused by a stopping layer for the heating particles, for the lower temperatures inferred for \jdbl and \joh (see the upper panel). 
As already presented in \citet[][see their Figure 1]{Bogdanov2021}, the difference in the beaming pattern between hydrogen and helium (or between fully  and partially ionized) atmosphere is typically smaller than 5\,\% at 1 keV or smaller photon energies. 
However, the relative difference starts to increase when the emission angle is larger than about $60\degr-70\degr$. 
At 1 keV, both partially ionized hydrogen and fully ionized helium show a similar monotonic decrease in emission at higher angles; fully ionized hydrogen, however, decreases more slowly and turns to limb brightening at the highest angles.  
We note that this difference is expected to be important only if both hot regions are observed at high emission angles. 
Otherwise, the radiation is dominated by the small angle emission, which has higher intensity. 
The effect could be larger in the case of \jdbl analysis without background estimates, since, there, only photons emitted with high emission angles can, typically, reach the observer, due to the inferred hot spot locations.
For example, emission angles are always above $57 \degr$ in the case of the maximum likelihood solution found for \texttt{ST-U} with a fully ionized hydrogen atmosphere. 

 \begin{deluxetable*}{cccccccc}
\tablewidth{0pt}
\tablecaption{Summary Table for \joh Runs}
\tablehead{\colhead{Model} & \colhead{Atmosphere} & \colhead{Free Beaming}  & \colhead{Background} & \colhead{Settings} & \colhead{$R_{\textrm{eq}}$ $[$km$]$} & \colhead{$M$ $[M_{\odot}]$} & \colhead{Evidence $\widehat{\ln\mathcal{Z}}$}}
\startdata
\texttt{ST-U}&
nsxHe &
No &
SW &
High &
$12.43_{-1.00}^{+1.19}$ $(13.23)$&
$2.071_{-0.065}^{+0.065}$ $(1.949)$ &
$-16160.31\pm0.07$
%-0.161603133692009942E+05  +/-    0.709310800303710742E-01
\\
\hline
\texttt{ST-U}&
nsxHp &
No &
SW &
High &
$12.40_{-0.97}^{+1.14}$ $(13.76)$&
$2.074_{-0.065}^{+0.065}$ $(2.047)$&
$-16160.83\pm0.07$
%-0.161608251525836567E+05  +/-    0.721225121921768297E-01
\\
\hline
\texttt{ST-U}&
nsxH &
No &
SW &
High &
$12.30_{-0.95}^{+1.16}$ $(12.82)$&
$2.072_{-0.066}^{+0.064}$ $(1.938)$&
$-16162.06\pm0.07$
%-0.161620647875850282E+05  +/-    0.712871001232151663E-01
\\
\hline
\texttt{ST-U}&
hatmH &
No &
SW &
High &
$12.41_{-0.99}^{+1.21}$ $(12.36)$&
$2.069_{-0.063}^{+0.065}$ $(2.014)$&
$-16162.07\pm0.07$
%-0.161620726350843706E+05  +/-    0.719896245410179547E-01
\\
\hline
\texttt{ST-U}&
nsxH &
max $10 \%$ &
SW &
High &
$12.18_{-0.88}^{+1.02}$ $(12.18)$&
$2.073_{-0.063}^{+0.064}$ $(1.994)$&
$-16162.27\pm0.07$
% -0.161622713901881161E+05  +/-    0.714743906610433993E-01
\\
\hline
\texttt{ST-U}&
nsxCp &
No &
SW &
High &
$11.93_{-0.72}^{+0.90}$ $(12.42)$&
$2.075_{-0.067}^{+0.066}$ $(2.145)$&
$-16174.34\pm0.08$
%-0.161743375101878664E+05  +/-    0.752382202111141485E-01
\\
\enddata
\tablecomments{\ \ For explanations of the atmosphere tables, see Table \ref{table:num_atmos}. 
The ``Free Beaming" column shows, if a beaming correction is allowed, how large the correction can be and whether it is energy-dependent ({\it E}-dep). 
The "Background" column shows whether background constraints have been applied using either \xmm (XMM) or the \NICER space weather estimate (SW), or if no constraints were applied (No). 
The ``Settings" column shows whether high- or low-resolution settings are used for the \XPSI hot region cells, phase bins in the star frame, and energy bins for specific photon flux (see \citealt{vinciguerra2023sim}), and if the multimodal (MM) variant of \MultiNest was used.
For all runs, we used $4\times10^{3}$ live points and $0.1$ sampling efficiency.
The $R_{\textrm{eq}}$ and $M$ columns show the radius and mass $68.3\,\%$ credible intervals around the median values and the values corresponding to the maximum likelihood sample (in parentheses).
}
\end{deluxetable*}\label{table:run_summary_2}

\begin{deluxetable*}{cccccccc}
\tablewidth{0pt}
\tablecaption{Summary Table for \jdbl Runs}
\tablehead{\colhead{Model} & \colhead{Atmosphere} & \colhead{Free Beaming} & \colhead{Background} & \colhead{Settings} & \colhead{$R_{\textrm{eq}}$ $[$km$]$} & \colhead{$M$ $[M_{\odot}]$} & \colhead{Evidence $\widehat{\ln\mathcal{Z}}$}} 
\startdata
\texttt{ST+PST} &
nsxHe &
No &
No &
Low-MM &
$15.39_{-0.79}^{+0.43}$ $(15.61)$&
$1.83_{-0.17}^{+0.13}$ $(1.97)$&
$-35778.17\pm0.07$
%-0.357781667343113950E+05  +/-    0.683236508337834725E-01
\\
\hline
\texttt{ST+PST} &
nsxH &
No &
No &
Low-MM &
$13.11_{-1.30}^{+1.30}$ $(13.09)$&
$1.37_{-0.17}^{+0.17}$ $(1.38)$&
$-35778.42\pm0.07$
%-0.357784187888312663E+05  +/-    0.701337417093998022E-01
\\
\hline
\texttt{ST-U} &
nsxHe &
{\it E}-dep, max $5 \%$ &
No &
High &
$15.44_{-0.75}^{+0.41}$ $(15.98)$&
%$14.46_{-1.61}^{+1.05}$ $(15.99)$&
$1.78_{-0.19}^{+0.16}$ $(2.00)$& 
%$1.50_{-0.21}^{+0.20}$ $(1.64)$&
$-35783.17\pm0.07$
%-0.357831704479664040E+05  +/-    0.674282015150491909E-01
%-0.357850071818225188E+05  +/-    0.665059140824213607E-01
\\
\hline
\texttt{ST-U} &
nsxH &
{\it E}-dep, max $5 \%$ &
No &
High &
$10.55_{-0.96}^{+1.28}$ $(10.02)$&
$1.14_{-0.10}^{+0.16}$ $(1.18)$&
$-35784.92\pm0.07$
%-0.357849180023939043E+05  +/-    0.663831880931019108E-01
\\
\hline
\texttt{ST-U} &
nsxHe &
No &
No &
High &
$14.00_{-1.88}^{+1.39}$ $(14.89)$&
$1.45_{-0.22}^{+0.21}$ $(1.70)$&
$-35785.22\pm0.07$
%-0.357852182064566805E+05  +/-    0.667224340725581511E-01
\\
\hline
\texttt{ST-U} &
nsxH &
No &
No &
High &
$10.59_{-0.92}^{+1.29}$ $(9.99)$&
$1.12_{-0.09}^{+0.15}$ $(1.05)$&
$-35786.21\pm0.07$ 
%-0.357862140619517158E+05  +/-    0.670418370039679162E-01
 \\
\hline
\texttt{ST-U} &
hatmH &
No &
No &
High &
$10.68_{-1.00}^{+1.32}$ $(11.08)$&
%$1.135_{-0.094}^{+0.152}$ $(1.181)$&
$1.14_{-0.09}^{+0.15}$ $(1.18)$&
$-35787.01\pm0.07$
%-0.357870145618711613E+05  +/-    0.671798098390887360E-01
\\
\hline
\texttt{ST-U} &
nsxHp &
No &
No &
High &
$12.94_{-0.81}^{+0.98}$ $(12.19)$&
$1.40_{-0.14}^{+0.15}$ $(1.40)$&
$-35796.14\pm0.07$
% -0.357961391789634654E+05  +/-    0.735673514638385539E-01
\\
\hline
\texttt{ST+PST$^{\mathrm{a}}$} &
nsxHp &
No &
No &
Low-MM &
$11.18_{-0.80}^{+0.88}$ $(12.15)$&
$1.22_{-0.12}^{+0.14}$ $(1.17)$&
$-35821.04\pm0.07$
%-0.358210408655789724E+05  +/-    0.610971332646975326E-01
\\
\hline
\texttt{ST-U} &
nsxCp &
No &
No &
High &
$10.95_{-0.65}^{+0.77}$ $(11.24)$&
%$1.030_{-0.022}^{+0.043}$ $(1.005)$& %1.0047
$1.03_{-0.02}^{+0.04}$ $(1.00)$&
$-35963.05\pm0.07$
%-0.359630479462840085E+05  +/-    0.721293503367116534E-01
\\
\hline
%\hline
\texttt{ST-U} &
nsxHe &
No &
XMM &
High-MM &
$15.22_{-1.10}^{+0.57}$ $(15.82)$&
$1.83_{-0.26}^{+0.23}$ $(2.11)$&
$-42694.06\pm0.07$
%-0.426940648536913504E+05  +/-    0.650333211501666619E-01
\\
\hline
\texttt{ST-U} &
nsxH &
No &
XMM &
High-MM &
$15.11_{-1.32}^{+0.65}$ $(15.19)$&
$1.88_{-0.20}^{+0.13}$ $(1.81)$&
$-42713.88\pm0.07$
%-0.427138792311895231E+05  +/-    0.667061219461037580E-01
\\
\hline
\texttt{ST-U} &
nsxHp &
No &
XMM &
High-MM &
$11.02_{-1.12}^{+1.21}$ $(10.97)$&
$1.27_{-0.14}^{+0.16}$ $(1.16)$&
$-42728.79\pm0.07$
% -0.427287890246201714E+05  +/-    0.668083021935379495E-01
\\
\enddata
\tablecomments{\ See the explanation for columns and their contents in Table \ref{table:run_summary_2}. 
Unlike the previous table, here, all the runs were performed with $1\times10^{4}$ live points and $0.3$ sampling efficiency. 
Note that the evidences for models with different background constraints are not comparable. \\
$^{\mathrm{a}}$ In addition to the worse evidence, the maximum likelihood found for this model is significantly worse than the likelihood obtained by plugging the best-fit \texttt{ST-U} parameter vector (from the run with nsxHp atmosphere) into the \texttt{ST+PST} framework ($-35769.98$ instead of $-35737.93$ in $\ln$ units).
This means that the sampler has not fully explored the parameter space.
}
\end{deluxetable*}\label{table:run_summary}

\section{Results}\label{sec:results}

In this section, we show the inference results for both \joh and \jdbl, when using different atmospheric choices. 
The main results, applying all the atmosphere tables, are shown in Figure \ref{fig:posterior_spacetime_num_J0740} for \joh and in Figure \ref{fig:posterior_spacetime_num_J0030} for \jdbl. 
The cases impacting the results the most (for \jdbl) are shown in Figures \ref{fig:posterior_spacetime_J0030_beam}--\ref{fig:posterior_spacetime_num_J0030_NxX}.
A summary of all the \joh runs is presented in Table \ref{table:run_summary_2}, and a summary of all the \jdbl runs is presented in Table \ref{table:run_summary}. 
In both cases, the rows are ordered based on decreasing evidence. 

\subsection{Mass and Radius Constraints for \joh with Different Atmospheres}\label{sec:param_all_atmos_J0740}

{
\begin{figure*}[t!]
\centering
\resizebox{\hsize}{!}{\includegraphics[
width=\textwidth]{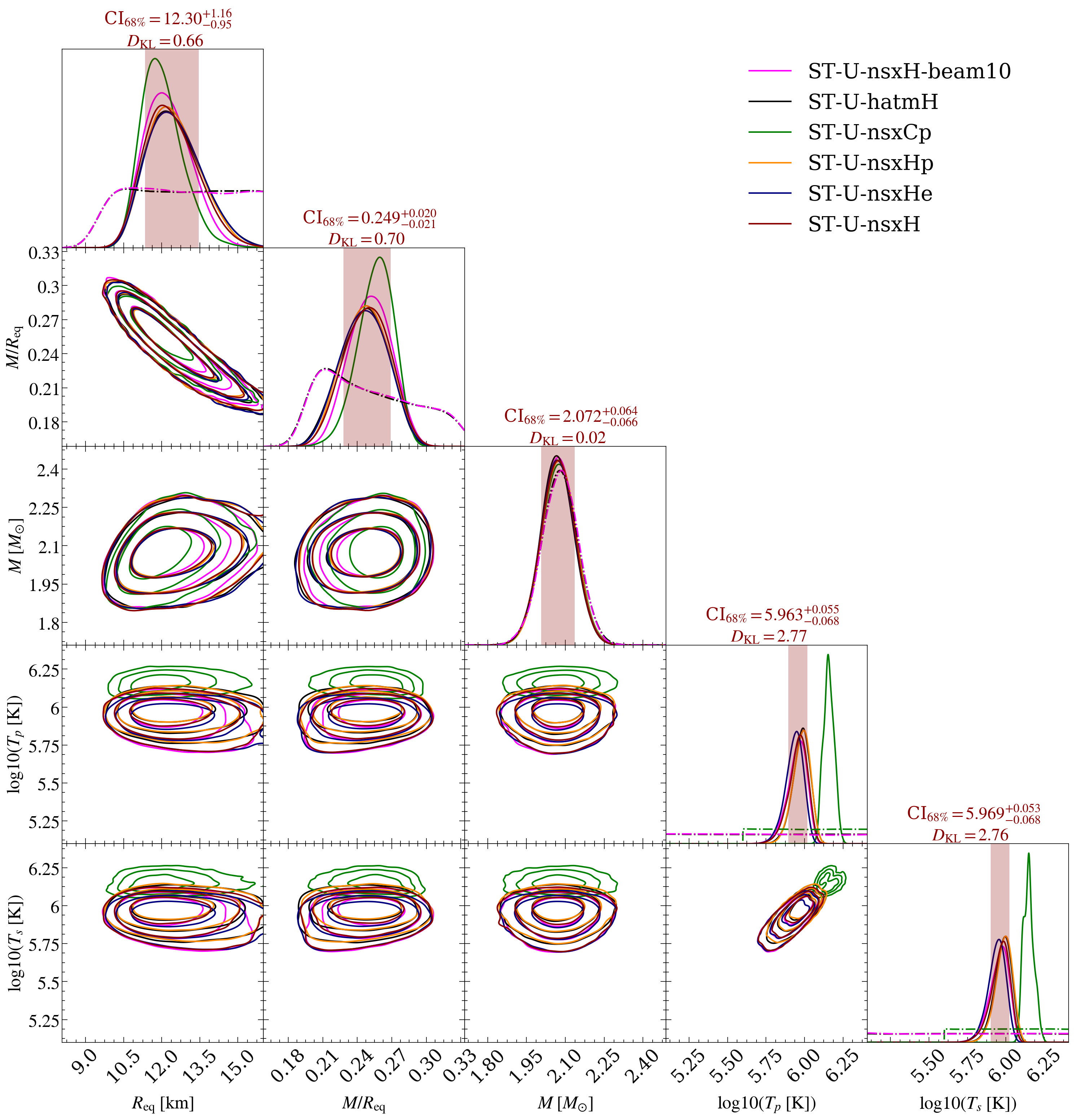}}
\caption{\small{
    Effect of the atmosphere model on the posterior distributions of radius, compactness, mass, and temperatures for both primary and secondary hot regions, using the \joh \NICER data set conditional on the \texttt{ST-U} model.  
    Six types of posterior distribution are shown with five different atmosphere tables used in the analysis (see Table \ref{table:num_atmos} for definitions), and one fully ionized hydrogen \texttt{NSX} model with a free beaming correction limited to 10\,\% deviation at most (ST-U-nsxH-beam10). 
    The credible intervals and the Kullback--Leibler (KL) divergence $D_{\mathrm{KL}}$ are reported for the fully ionized hydrogen \texttt{NSX} case: ST-U-nsxH (see Table \ref{table:run_summary_2} for mass and radius values in other cases).  
    The prior distributions are shown by the dashed--dotted functions.  
    The 1D intervals (shaded in red) contain $68.3\,\%$ of the posterior mass, and the contours in the 2D panels contain $68.3\,\%$, $95.4\,\%$, and $99.7\,\%$ of the posterior mass.     
    For more details of the figure elements, see Figure 5 of \citet{Riley2021} and Figure 5 of \citet{salmi2022}. 
    Posterior distributions for the other parameters are shown in Figure \ref{fig:posterior_num_J0740_appendix} of Appendix \ref{sec:appendix}.  
}}
\label{fig:posterior_spacetime_num_J0740}
\end{figure*}
}

We start by presenting the inference results for \joh
 with different numerical atmospheres
and using the \texttt{ST-U} hot region model in Figure \ref{fig:posterior_spacetime_num_J0740} (see
also the full list of inferred radius, mass, and evidence values in Table \ref{table:run_summary_2}). 
As mentioned in Section \ref{sec:xpsi_modeling}, we also use the  space weather background estimate as a lower limit for the \NICER background.
We find no significant differences in the inferred NS radii between any of the atmosphere models. 
The only model for which the posterior contours can be clearly distinguished from the rest is the partially ionized carbon atmosphere (especially in the inferred temperatures). 
However, as in the case of \jdbl (see Section \ref{sec:STU_NICER_only_J0030_num}), the carbon model is significantly disfavored based on the evidence values (more than $\sim 10$ difference in natural logarithmic, i.e., ln units). 
The evidence differences between the other atmosphere cases are only marginal. 
The results also show that allowing 10\,\% maximum deviation in the beaming pattern using the energy-independent beaming parameters (as explained in Section \ref{sec:beaming_formalism}) for fully ionized hydrogen does not have any significant impact on the results.

{
\begin{figure*}[t!]
\centering
\resizebox{\hsize}{!}{\includegraphics[
width=\textwidth]{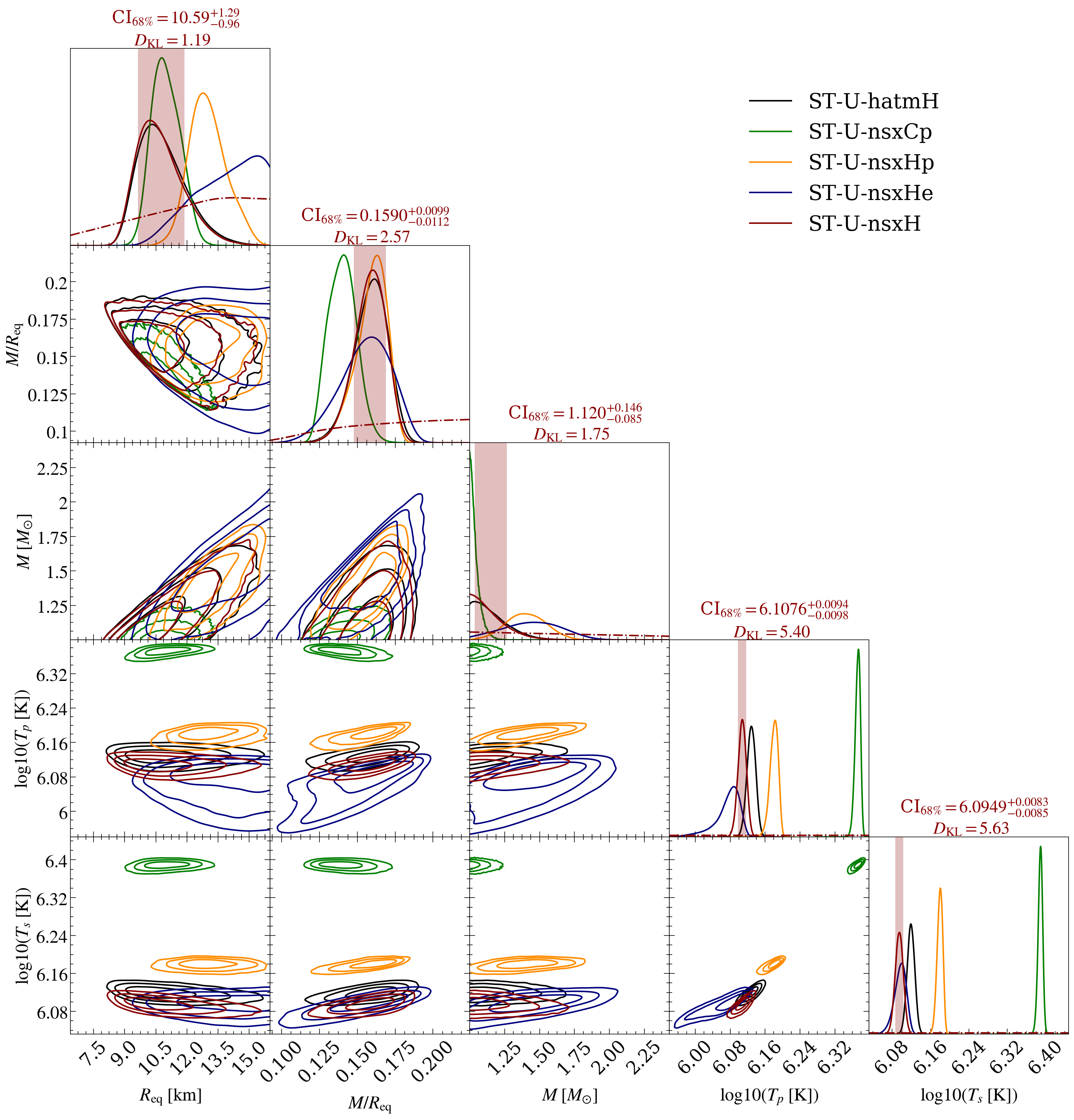}}
\caption{\small{
    Effect of the atmosphere model on the posterior distributions of radius, compactness, mass, and temperatures for both primary and secondary hot regions, using the \jdbl \NICER data set conditional on the \texttt{ST-U} model. 
    Five types of posterior distribution are shown with different atmosphere tables used in the analysis (see Table \ref{table:num_atmos} for definitions). 
    The 1D credible intervals and the KL-divergence estimates are reported for the fully ionized hydrogen \texttt{NSX} case: ST-U-nsxH (see Table \ref{table:run_summary} for mass and radius values in other cases). 
    See the caption of Figure \ref{fig:posterior_spacetime_num_J0740} for additional details about the figure elements. 
    Posterior distributions for the other parameters are shown in Figure \ref{fig:posterior_num_J0030_appendix} of Appendix \ref{sec:appendix}.
}}
\label{fig:posterior_spacetime_num_J0030}
\end{figure*}
}

\subsection{Mass and Radius Constraints for \jdbl with Different Atmospheres}\label{sec:J0030}

\subsubsection{ST-U NICER-only Fit}\label{sec:STU_NICER_only_J0030_num}

Next, we present the inference results for \jdbl with different numerical atmospheres and using the \texttt{ST-U} hot region model in Figure \ref{fig:posterior_spacetime_num_J0030} (see also Table \ref{table:run_summary}). 
We see that the radius inferred using the externally heated, fully ionized hydrogen atmosphere model is consistent with the one from the corresponding deep-heated \texttt{NSX} model. 
There is only a small shift (smaller than $\sim 0.03$ in $\log_{10} T [\mathrm{K}]$) in the inferred temperature for both hot regions to higher values when using the externally heated model. 
This is expected because the spectrum peaks at slightly lower energy than the corresponding deep-heated model with the same model parameters (see Figure \ref{fig:spectrum}). 
However, at least in this temperature range and with the assumed return-current properties mentioned in Section \ref{sec:num_atmos}, this effect is too small to cause any significant change in the inferred radius. 

On the contrary, both the partial ionization and the chemical composition have significant effects on the inferred radius. 
The use of a partially ionized hydrogen model instead of a fully ionized atmosphere shifts the median radius and its 68.3\,\% credible interval from $10.59_{-0.92}^{+1.29}$ to $12.94_{-0.81}^{+0.98}$ km, i.e., leaving no overlap in the 68.3\,\% intervals. 
The inferred median mass also increases from around $1.1$ to $1.4$ \msol, and the temperatures of both hot regions become higher (again as expected based on the spectral shapes in Figure \ref{fig:spectrum}).
As seen from Figure \ref{fig:posterior_num_J0030_appendix} of Appendix \ref{sec:appendix}, there are also small changes in the other model parameters. 
However, both hot regions are still inferred to be on the same hemisphere and highly inclined toward the observer. 
The evidence for the model with partially ionized hydrogen is worse than for the model with fully ionized hydrogen, by 10 ln units (see Table \ref{table:run_summary}).

Changes in the chemical composition of the atmosphere also cause the inferred radius to change, in the case of \jdbl.
For partially ionized carbon, the inferred radius 68.3\,\% credible interval is $10.95_{-0.65}^{+0.77}$ km, which does not overlap with the corresponding interval of partially ionized hydrogen. 
In addition, major differences are seen in many of the other parameters.
However, based on the evidence, the differences compared to the other models, which are more than $\sim 200$ in ln units (see Table \ref{table:run_summary}), and the quality of the residuals between the data and best-fit model, the carbon model is clearly disfavored for \jdbl. 
For fully ionized helium, the evidence is similar (or even slightly higher) to that of hydrogen, and the effect on the shift in the inferred radius is the largest; from $10.59_{-0.92}^{+1.29}$ (for hydrogen) to $14.00_{-1.88}^{+1.39}$ km (for helium). 
The inferred mass also increases to around $1.5$ \msol. 
Most of the other model parameters are not greatly affected, {although there is more posterior mass for the hot regions and observer inclination to be closer to the equatorial plane compared to the hydrogen case. 

{
\begin{figure*}[t!]
\centering
\resizebox{\hsize}{!}{\includegraphics[
width=\textwidth]{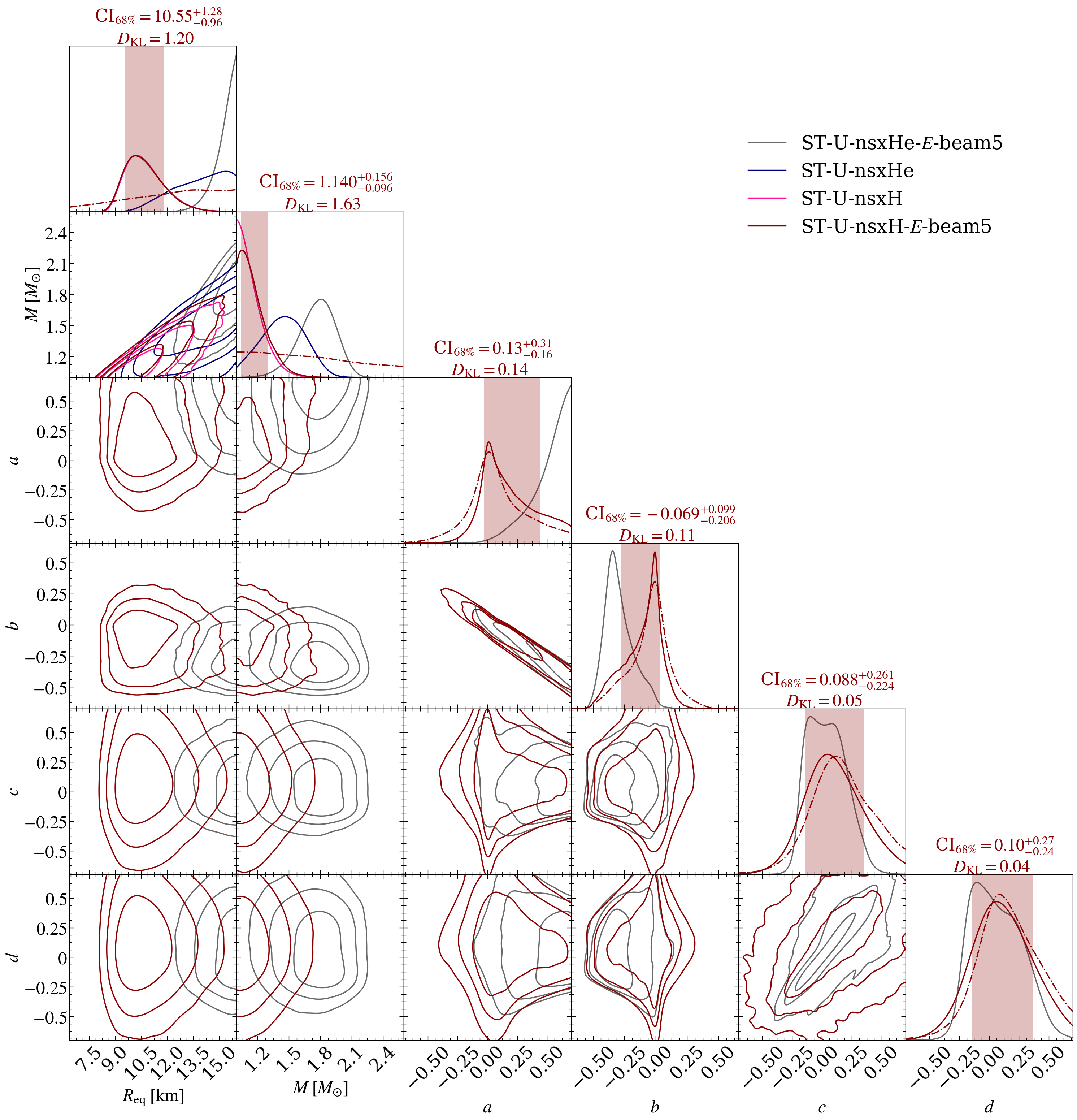}}
\caption{\small{
    Effect of the atmosphere model on the posterior distributions of radius, mass, and beaming parameters $a$, $b$, $c$, and $d$, using the \jdbl \NICER data set conditional on the \texttt{ST-U} model and energy-dependent beaming parameterization. 
    Priors (dashed--dotted curves) and posteriors for the four beaming parameters are shown only for the models where those parameters were kept free. 
    The same two hydrogen and helium posteriors without beaming modifications are also shown, as in Figure \ref{fig:posterior_spacetime_num_J0030} (ST-U-nsxH and ST-U-nsxHe), in the radius and mass panels. 
    The posteriors for ST-U-nsxH and ST-U-nsxH-$\it{E}$-beam5 models are almost exactly overlapping. 
    The credible intervals and KL-divergence estimates are reported for the latter (see Table \ref{table:run_summary} for mass and radius values in other cases). 
    See the caption of Figure \ref{fig:posterior_spacetime_num_J0740} for additional details about the figure elements. 
    Posterior distributions for the other parameters are shown in Figure \ref{fig:posterior_beam_J0030_appendix} of Appendix \ref{sec:appendix}.
}}
\label{fig:posterior_spacetime_J0030_beam}
\end{figure*}
} 

\subsubsection{ST-U NICER-only Fit with Free Beaming}\label{sec:STU_NICER_only_J0030_beam}

As explained in Section \ref{sec:beaming_formalism}, we have also performed a few runs with additional beaming parameters introduced to add more freedom in the predicted beaming pattern. 
The main results of these runs, for fully ionized hydrogen and helium, are shown in Figure \ref{fig:posterior_spacetime_J0030_beam} and in Table \ref{table:run_summary}. 
We see that limiting the beaming correction to 5\,\% at maximum (at any energy or emission angles below $60\degr$) has no significant effects on the radius and mass constraints in the hydrogen case. 
For helium, the median radius shifts from 14.0 to 15.4 km, but the 68.3\,\% credible intervals are still overlapping. 
For hydrogen, no significant information gain from prior to posterior distribution was found in any of the beaming parameters. 
The inferred values thus peak around zero in all the correction factors $a$, $b$, $c$, and $d$, following closely the prior distribution. 
Also, the uncertainty in these parameters did not cause any noticeable broadening of the credible intervals of radius or mass. 
For helium, however, the inferred $a$ and $b$ differ significantly from zero (hitting the upper bound of 0.7 for $a$), and the radius and mass constraints become tighter most likely due to the cut in radius at 16 km.

{
\begin{figure*}[t!]
\centering
\resizebox{\hsize}{!}{\includegraphics[
width=\textwidth]{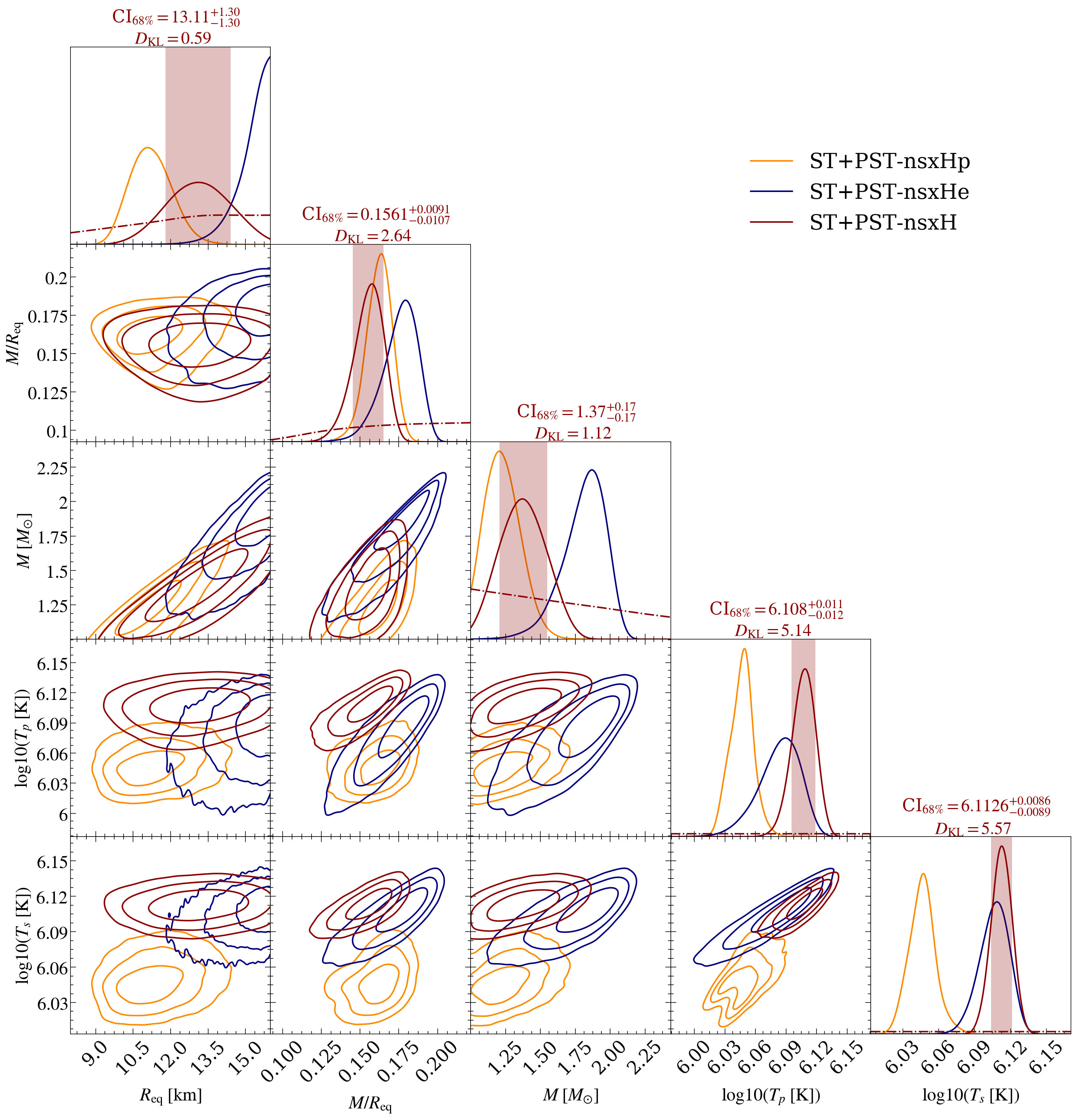}}
\caption{\small{
    Effect of the atmosphere model on the posterior distributions of radius, compactness, mass, radius, and temperatures for both primary and secondary hot regions, using the \jdbl \NICER data set conditional on the \texttt{ST+PST} model. 
    Posterior distributions are shown for fully ionized hydrogen and helium, and for partially ionized hydrogen \texttt{NSX} models. 
    The credible intervals and KL-divergence estimates are reported for the ST+PST-nsxH model (see Table \ref{table:run_summary} for mass and radius values the other models).
    See the caption of Figure \ref{fig:posterior_spacetime_num_J0740} for additional details about the figure elements. 
    Posterior distributions for the other parameters are shown in Figure \ref{fig:posterior_num_J0030_stpst_appendix} of Appendix \ref{sec:appendix}.  
}}
\label{fig:posterior_spacetime_num_J0030_stpst}
\end{figure*}
}

\subsubsection{ST+PST NICER-only Fit}\label{sec:STPST_NICER_only_J0030_num}

We also performed a few runs with the more complex and computationally expensive \texttt{ST+PST} model for the hot region shapes, as mentioned in Section \ref{sec:xpsi_modeling}. 
This model corresponds to that used for the headline results of \citet{RWB_nicer19}. 
Since this model is computationally more expensive, runs were only done for the fully ionized hydrogen and helium, and partially ionized hydrogen cases, in which the difference in radius was largest.
The results are shown in Figure \ref{fig:posterior_spacetime_num_J0030_stpst} and in Table \ref{table:run_summary}. 
We see that both the inferred radius and mass increase when using a helium instead of a hydrogen atmosphere (both fully ionized). The radius changes from $13.11_{-1.30}^{+1.30}$ to $15.39_{-0.79}^{+0.43}$ km (the latter being limited by the 16 km prior upper limit), and mass from $1.37_{-0.17}^{+0.17}$ to $1.83_{-0.17}^{+0.13}$ \msol (which is still smaller than the $~2.7$ \msol value found in \citealt{MLD_nicer19} when using a helium model; see discussion in Section \ref{sec:disc_composition}).
Again, there is no substantial difference in the evidence between helium and hydrogen.

In the case of partially ionized hydrogen, the inferred radius is $11.18_{-0.80}^{+0.88}$ km, which barely overlaps with the 68.3\,\% interval of fully ionized hydrogen. 
Significant differences are found in many of the other parameters, for example, inferring more antipodal-like geometry with hot regions and observer inclination close to the equatorial plane (as seen from Figure \ref{fig:posterior_num_J0030_stpst_appendix} of Appendix \ref{sec:appendix}).
However, the maximum likelihood found is more than 30 $\ln$ units worse than the likelihood for the corresponding best-fit \texttt{ST-U} solution in the \texttt{ST+PST} framework.
This means that even the high-resolution settings ($1\times10^{4}$ live points; and $0.3$ sampling efficiency) were not sufficient to fully explore the parameter space. 

As mentioned in Section \ref{sec:posterior_computation}, the mode-separation variant was used for these runs to better identify and explore multiple modes found by the sampler. 
Using that, a secondary mode with worse evidence was found for all the models.
The second mode for fully ionized hydrogen corresponds to a radius of around 11 km (instead of 13 km of the main mode), while the second modes of fully ionized helium and partially ionized hydrogen have similar radii to the corresponding main modes (i.e., 15 and 11 km, respectively). 

{
\begin{figure*}[t!]
\centering
\resizebox{\hsize}{!}{\includegraphics[
width=\textwidth]{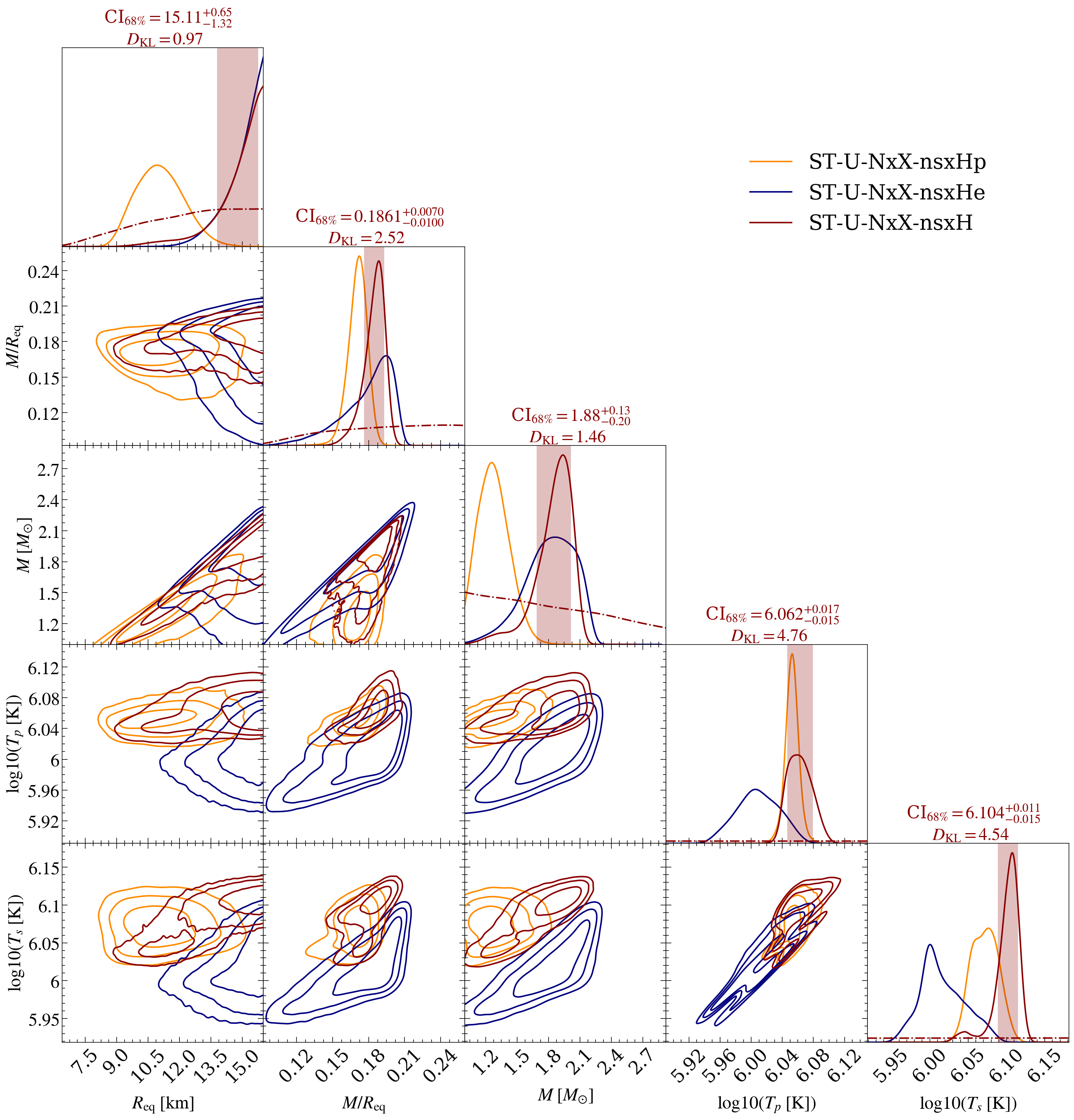}}
\caption{\small{
    Effect of the atmosphere model on the posterior distributions of radius, compactness, mass, and temperatures for both primary and secondary hot regions, using the \jdbl joint \NICER and \xmm (NxX) data sets conditional on the \texttt{ST-U} model.  
    Posterior distributions are shown for fully ionized hydrogen and helium, and partially ionized hydrogen \texttt{NSX} models.    
    The credible intervals and KL-divergence estimates are reported for the ST-U-NxX-nsxH model (see Table \ref{table:run_summary} for mass and radius values the other models).
    See the caption of Figure \ref{fig:posterior_spacetime_num_J0740} for additional details about the figure elements. 
    Posterior distributions for the other parameters are shown in Figure \ref{fig:posterior_num_J0030_NxX_appendix} of Appendix \ref{sec:appendix}.
}}
\label{fig:posterior_spacetime_num_J0030_NxX}
\end{figure*}
}

\subsubsection{ST-U \NICER and \xmm Fit}\label{sec:STU_NICER_XMM_J0030_num}

As in \citet{vinciguerra2023bravo}, we also performed a few runs where the \NICER background signal is constrained by fitting jointly \NICER and  \xmm data. 
The results are shown in Figure \ref{fig:posterior_spacetime_num_J0030_NxX} and in Table \ref{table:run_summary}. 
We see that the inferred radius is equally high for both the fully ionized hydrogen and helium models (unlike in the previous cases), i.e., close to the prior upper limit of 16 km.  The mass is around 1.9 \msol and the hot region geometry is similar (more antipodal compared to the corresponding runs without \xmm) in both the hydrogen and helium cases. 
However, a few other parameters, including the temperatures and interstellar absorption parameter $N_{\mathrm{H}}$, are inferred to have notably different credible intervals. 
Interestingly, the evidence for the helium case is this time significantly higher than for the hydrogen case (around 20 in ln units). 

Hot regions close to the equatorial plane are also inferred using the partially ionized hydrogen atmosphere.
In this case, however, the inferred radius is significantly smaller, around 11 km. 
Nevertheless, the evidence for the partially ionized hydrogen model is around 15 ln units worse than that for the fully ionized hydrogen.

The mode separation (see Section \ref{sec:posterior_computation}) allowed us to identify, in the case of the fully ionized hydrogen atmosphere, two distinct modes \citep[similarly to][]{vinciguerra2023bravo}.
The secondary mode corresponds to a smaller radius around 11 km and a slightly different geometry, but with a significantly smaller local evidence. 
In the case of the fully ionized helium atmosphere, we found four distinct modes, which all correspond to radii above 15 km with slightly different geometries and local evidences better than in the best mode found for the hydrogen run. In the case of the partially ionized hydrogen atmosphere, we found two modes, where the secondary mode corresponds to a radius of around 12~km instead of the 11~km of the primary mode.

\section{Discussion}\label{sec:discussion}

As shown in Section \ref{sec:results}, we found that the radius inferred for \jdbl is sensitive to some of the assumptions made for the atmosphere model. 
On the other hand, the results for \joh were shown to be largely insensitive to any of the model assumptions that we considered here. 
In the following, we discuss possible reasons for this difference and what can be learned about the atmosphere properties of these sources based on our study.

\subsection{Differences between J0030 and J0740}\label{sec:disc_J0030_J0740_diff}

There may be several reasons why the \joh analysis appears to be less affected by the atmosphere model choices than the \jdbl analysis. 
One of them is the quality of the data; due to the larger number of total counts in the \jdbl data set ($1.5\times10^{6}$ versus $0.6 \times10^{6}$) %($1518737$ vs $628280$)
and the higher background for \joh, the credible intervals are notably tighter for the former and more sensitive to systematic errors caused by unaccounted-for physics.

However, the sensitivity to atmosphere could also depend on the properties of the star itself. 
The radio observations of \joh constrain the observer’s inclination to be very close to the star’s orbital plane (and thus the star's equatorial plane). The highest likelihood solutions have a hot spot that also tends to lie close to the equatorial plane. As a result of the constrained geometry, the hot spot on the equator is brightest when facing the observer, which is also when the light is emitted normal to the surface, minimizing the effect of beaming by the different atmosphere models. 
The inclination of \jdbl is unconstrained by radio observations, and the highest likelihood solutions tend to favor, in most cases, models where the spots are close to the limb of the star, as viewed by the observer. As a result, the emitted light that reaches the observer tends to be emitted close to the tangent to the surface, where differences in beaming are much more pronounced. 
This can result in a stronger dependence on the atmosphere model for \jdbl.

As presented in Section \ref{sec:STU_NICER_XMM_J0030_num}, the atmosphere effects, for hydrogen versus helium, on the inferred radii for \jdbl are still small when we consider the joint \NICER and \xmm runs. 
The similarity could be caused either by the hard cut in the prior upper limit at 16 km or by the fact that the inferred results for both hydrogen (see also \citealt{vinciguerra2023bravo}) and helium predict significantly different hot region geometries than those without background constraints.
In this case, most of the detected photons are emitted at smaller emission angles (down to around $5\degr$ instead of $60\degr$) where the relative difference between the atmosphere models is smaller. 
However, the inferred radius for the corresponding partially ionized hydrogen case is significantly smaller (around 11 km), even with a similar best-fit geometry where the detected photons are dominated by the small angle emission.

Besides the data quality and viewing geometry, it could also be that external parameter constraints help to reduce the free parameter space such that different results with different atmosphere models are not found.
Thus, the tight priors on mass, distance, and inclination for \joh could also explain its insensitivity to the atmosphere model choices.

\subsection{Implications for Chemical Composition}\label{sec:disc_composition}

In the case of both \jdbl and \joh, both hydrogen and helium compositions of the atmosphere were found to mostly fit the data equally well. 
However, the inferred radii of \jdbl using helium tend to hit the 16 km upper limit of the prior, making these solutions less plausible.  Differences in the evidence are only a few units in natural logarithm (with helium always being better), except in the case of the joint \NICER and \xmm analysis on \jdbl, where the evidence for helium is around 20 units larger than for hydrogen. 
However, it could still be possible that small differences in the constrained background might lead to substantially different NS parameters and preferred atmosphere. 
It would therefore be beneficial to perform ancillary analysis using direct background estimates for \NICER as in \citet{salmi2022}, which would give information about the background independently from \xmm. 

When comparing the evidences between hydrogen and helium, one could also weigh the evidence based on how much more likely the hydrogen composition is considered a priori. 
However, quantifying this is difficult. 
It depends on the probability that the NS has accreted from a hydrogen-poor companion and/or that the diffusive nuclear burning has converted all hydrogen to helium. 
In addition, if even a small amount of hydrogen \citep[$\sim 10^{-19}$ \msol,][]{Romani1987} has been accreted after the mass transfer from the companion ended, it would be enough to create an upper atmosphere of pure hydrogen \citep{Blaes1992,Wijngaarden2019,Wijngaarden2020,Bogdanov2021}. 

As noted in Section \ref{sec:STPST_NICER_only_J0030_num}, our helium results for \jdbl differ from those in \citet{MLD_nicer19}; we infer a significantly lower NS mass, $1.8$ \msol (for \texttt{ST+PST}), rather than $2.7$ \msol. 
This can likely be explained by the different prior assumptions, especially the more strict 16 km upper limit for the radius in the analysis of this paper, since this also limits how high a mass can be obtained for a given NS compactness (see Figure \ref{fig:posterior_spacetime_num_J0030_stpst}).
In addition, \citet{MLD_nicer19} uses a completely independent analysis pipeline with the differences discussed, e.g., in \citet{Miller2021}, \citet{Riley2021}.

As mentioned in Section \ref{sec:intro}, a composition with elements heavier than helium might also be possible, although less likely. 
Given that we found significantly worse evidences for the tested carbon atmosphere cases (more than $\sim 200$ for \jdbl and more than $\sim 10$ for \joh in ln-units), it seems clear that carbon atmospheres are very unlikely for these sources. 
Having any other heavy element composition or a mix of different elements was not considered in this study, as they are both considered unlikely. In particular the latter seems improbable due to the rapid sinking of elements via diffusive gravitational separation as explained in Section \ref{sec:intro}.

\subsection{Implications for Partial Ionization}\label{sec:disc_ionization}

Some of the atmosphere models we applied allowed the possibility of the atoms in the atmosphere to be partially ionized. 
The effect of ionization state is usually expected to be highest for colder atmospheres where the ionization fraction is noticeably smaller than unity. 
However, since the atmosphere consists of layers with different temperatures, even an NS with a relatively high effective temperature might have shallow layers with low enough temperatures to produce bound--bound and bound--free opacity features in the escaping radiation. 
This could explain why the inferred radius results for \jdbl (see Section \ref{sec:STU_NICER_only_J0030_num}) appear to be sensitive to the choice of ionization state, even if the inferred effective temperature is around $\log_{10} (T_{\mathrm{eff}}/\mathrm{K}) \approx 6.1$ for both hot regions. 

Nonetheless, at higher temperatures, the partially ionized atmosphere calculation becomes more inaccurate due to the opacity uncertainties for a range of photon energies \citep{IglesiasRogers1996,Badnell2005,Colgan2016,Bogdanov2021}. 
Thus, interpreting the partially ionized results demands some extra caution. 
For $\log_{10} (T_{\mathrm{eff}}/\mathrm{K}) \approx 6.1$, the opacity uncertainties should still be small, but larger at $\log_{10} (T_{\mathrm{eff}}/\mathrm{K}) \approx 6.4$, which is inferred for both hot regions of \jdbl with the partially ionized carbon case. 
Despite these uncertainties, we count the much worse evidence of the carbon model as a strong indication that the atmosphere of \jdbl is not composed of carbon. 

In the case of \joh, we found a good agreement between the fully and partially ionized hydrogen results, which is consistent with that from \citet{Miller2021}, \citet{Riley2021}, \citet{salmi2022}, where the code applying the partially ionized table \citep{Miller2021} obtained consistent results with those using the fully ionized table \citep{Riley2021}. 
In our results, the evidence difference between the two models is not significant (partially ionized model performs better by less than 2 in ln units).
In the case of \jdbl, however, we always find significantly worse evidence when using the partially ionized hydrogen model (at least 10 ln units in all cases). 
This suggests that the fully ionized models should be preferred over the partially ionized models for \jdbl; however, the latter predicts more reasonable NS radii when including \xmm data (11 instead of 15 km) for the \texttt{ST-U} spot configuration.

\subsection{Other Atmospheric Effects}\label{sec:disc_other}

In addition to chemical composition and ionization state of the atmospheric elements, some other atmosphere model effects on the NS radius constraints were also considered in Section \ref{sec:results}. 
One of them is the stopping depth of the return-current particles in the atmosphere. 
We studied this by using an externally heated, fully ionized hydrogen atmosphere model with certain assumptions in the properties of the return-current particle energy distribution (see Section \ref{sec:num_atmos}). 
In this case, no significant changes in the inferred radius were observed for either \jdbl or \joh when compared to the corresponding deep-heated model. 
As mentioned in Section \ref{sec:STU_NICER_only_J0030_num}, this shows that even the slow bombarding particles (with average Lorentz factors of $100$) do not stop at depths shallow enough to affect NS radius constraints. 
In principle, the effects could be larger for NSs with hotter spots (e.g., around $\log_{10} T_{\mathrm{eff}} = 6.4$), where the impact on the emergent intensity is larger, as seen in Section \ref{sec:model_comparison}. 
However, even then, the effect is likely to be unimportant, since the return current particles are more likely notably faster than those assumed here, based on pulsar magnetosphere simulations \citep{HardingMuslimov2011,TA13,Chen2014,Cerutti2017,Philippov2018}.
Most of them have focused on slowly rotating and highly magnetic pulsars, but the particle energies are expected to be larger for the millisecond pulsars with lower magnetic fields (see Figure 10 of \citealt{HardingMuslimov2011} and the discussion in \citealt{Bogdanov2021}).

We also verified that the numerical implementation of the atmosphere calculation is unlikely to have large effects on the NS parameter constraints since the externally heated atmosphere table was produced with a completely different algorithm. 
It seems that the smaller grid resolution of that atmosphere table (see Table \ref{table:num_atmos} for a summary) does not affect the results either. 
However, there are still effects that we have not considered in this study.
One of them is the effect of the magnetic field strength on the atmosphere. 
As discussed in \citet{Bogdanov2021}, such effects on the spectrum and beaming pattern should be negligible if \jdbl has a purely dipolar magnetic field. 
In the presence of multipoles, the field strength could get higher, but the NS atmosphere would still likely be not significantly affected by the magnetic field \citep{Kalapotharakos2021}.  
Another possibility with a high magnetic field (unlikely for RMPs) would be the formation of a bare condensed NS surface, instead of an atmosphere \citep{Medin2007}.

\subsection{Beaming Uncertainty}\label{sec:disc_beaming}

Ideally, the uncertainty in the atmosphere model assumptions and numerical calculations could be parameterized (as attempted in Section \ref{sec:beaming_formalism}) instead of performing multiple runs with different assumptions. 
Our empirical beaming correction function allowed reasonable freedom in the predicted angular dependency of radiation. 
However, the NS radius results were shown to be largely not sensitive to the allowed beaming correction in all the cases we considered (see Sections \ref{sec:STU_NICER_only_J0030_beam} and \ref{sec:param_all_atmos_J0740}), which is in contrast to the much higher sensitivity found when applying different atmosphere tables for \jdbl. 
In the case of hydrogen, the values for all the beaming correction factors $a$, $b$, $c$, and $d$ were found to follow closely their prior distributions, centered around zero. 
In the case of helium, significantly different values were inferred for $a$ and $b$, but the nominal beaming was still changed only by $6\,\%$ on average at 1 keV, while the corresponding average difference between the nsxH and nsxHe tables is about 2 times larger.

There are many reasons that could explain the smaller sensitivity to the beaming correction compared to the atmosphere table choice. 
One is that allowing a change in only the beaming pattern may not be enough, and we should also account for the change in the emergent spectrum, if we want to allow similar parameter space to be explored when using either a beaming-modified hydrogen or helium atmosphere. 
Another reason could be that we allowed no more than a 5\,\% deviation in beaming in our \jdbl runs at emission angles below $60\degr$, which can indirectly limit the correction also at highest angles (due to the functional form of the beaming correction) to be less than the differences between the atmosphere tables (which rapidly increase at the highest angles for some cases).
An additional reason could be that the form of the correction function is not general enough, to accurately capture the differences between the different atmosphere models even at smaller emission angles. 
A more detailed analysis of this is left for future work.

\subsection{Comparison to Other Modeling Uncertainties}\label{sec:comparison_uncertanties}

Based on our results, and discussion in Section \ref{sec:disc_J0030_J0740_diff}, it seems clear that the sensitivity of the mass and radius constraints to NS atmospheric effects can vary a lot from star to star. 
Due to the complicated degeneracies between atmospheric beaming, hot spot geometries, and background radiation, it is difficult to determine which of these factors is most important.
Different assumptions in the atmosphere can lead to different modes being preferred in the multimodal posterior surface, especially when the solutions (e.g., in terms of geometry parameters) favored in some atmosphere cases are pushed against the radius or mass prior limits in other atmosphere cases. 
In this case, there might be no clear trend between two atmosphere models for the same star, as seen in the fully versus partially ionized hydrogen comparison for \jdbl, where partial ionization results in significantly smaller or higher NS radii than those of full ionization depending on the background assumption.
\footnote{It should be also noted that the adequate sampling of the parameter space has not been rigorously proved for all the different atmospheres, hot spot shapes, and background cases. 
} 
On the positive side, the sensitivity to the atmosphere models allows us also to discriminate between the models with lower evidence or with unexpected values for the inferred parameters. 
However, the favored atmosphere models and the inferred values may not necessarily remain the same when external constraints or more complex spot patterns are implemented.

\section{Conclusions}\label{sec:conclusions}

We have explored the sensitivity of the constraints that we obtain on NS parameters, especially the radius, to the assumptions made for the NS atmosphere. 
We found that none of our cases had any significant impact on the inferred radius and other properties of \joh, possibly because of its relatively low signal-to-noise ratio, its inferred viewing geometry, or its external parameter constraints. 
However, results for \jdbl were found to be potentially sensitive to both the chemical composition of the atmosphere and the choice of whether partial ionization in the atmosphere was accounted for or not. 
The largest difference would appear if the atmosphere were to consist of helium instead of hydrogen, which would shift the radius from around 10.5 to 14 km in the case of the \texttt{ST-U} hot region model (circular hot regions), or from 13 to 15 km in the case of \texttt{ST+PST} hot region model (a circle and a crescent).  
In both cases, the difference in the evidence is not enough to distinguish between the models, although hydrogen composition is expected to be a priori more likely based on evolutionary arguments. 

When constraining the \jdbl background using \xmm \citep[as in][]{vinciguerra2023bravo} with the \texttt{ST-U} model, the inferred radii using fully ionized hydrogen and helium are consistent with each other but close to the prior upper limit of 16 km. 
This consistency could be, at least partly, caused by the different inferred hot region geometry, where more radiation is emitted at smaller emission angles, for which the fractional differences between the atmosphere models are smaller. 
However, the corresponding partially ionized hydrogen case predicts a significantly smaller NS radius (around 11 km) even with a similar hot region geometry. 
Thus, more studies with accurate background and other constraints may be needed to properly resolve the typically multimodal structure of the posterior surface for \jdbl. 

In addition, we also found atmosphere cases that either did not significantly impact the \jdbl and \joh results or were shown to be unlikely based on the inferred evidence. 
The former includes the runs with an externally heated model with certain assumptions for the return-current energy distribution, and runs with parameterized uncertainty in atmospheric beaming pattern. %limited to a maximum 5\,\% deviation for the \jdbl analysis and a maximum 10\,\% deviation for the \joh analysis. 
The latter includes the runs with carbon composition, which does not fit the data as well as the other models. 

\begin{acknowledgments}
This work was supported in part by NASA through the \NICER mission and the Astrophysics Explorers Program.
T.S., S.V., D.C., A.L.W., Y.K., and B.D. acknowledge support from European Research Council (ERC) Consolidator Grant (CoG) No.~865768 AEONS (PI: Watts). 
%This work was sponsored by NWO Domain Science for the use of the national computer facilities. 
This work used the Dutch national e-infrastructure with the support of the SURF Cooperative using grant No. EINF-4664. 
Part of the work was carried out on the HELIOS cluster including dedicated nodes funded via the above-mentioned ERC CoG. 
The computer resources of the Finnish IT Center for Science (CSC) and the FGCI project (Finland) are also acknowledged for part of the new atmosphere grid computation. 
W.C.G.H. appreciates the use of computer facilities at the Kavli Institute for Particle Astrophysics and Cosmology and acknowledges support through grant 80NSSC22K0397 from NASA.
S.G. acknowledges the support of the Centre National d'Etudes Spatiales (CNES). S.M.M. thanks NSERC for its support.
T.S. would like to thank Juri Poutanen, Thomas Riley, Nathalie Degenaar, Cole Miller, Joonas Nättilä, and the anonymous referee for useful discussions and comments. 
\end{acknowledgments}

\bibliographystyle{aasjournal}
\bibliography{allbib}

\clearpage 
\appendix
\section{Additional Figures}
\label{sec:appendix}

In Figures \ref{fig:posterior_num_J0740_appendix}--\ref{fig:posterior_num_J0030_NxX_appendix}, we show the posterior distributions for the NS radii together with other model parameters, not including those presented in the Figures \ref{fig:posterior_spacetime_num_J0740}--\ref{fig:posterior_spacetime_num_J0030_NxX} of the main text. 

{
\begin{figure*}[t!]
\centering
\resizebox{\hsize}{!}{\includegraphics[
width=\textwidth]{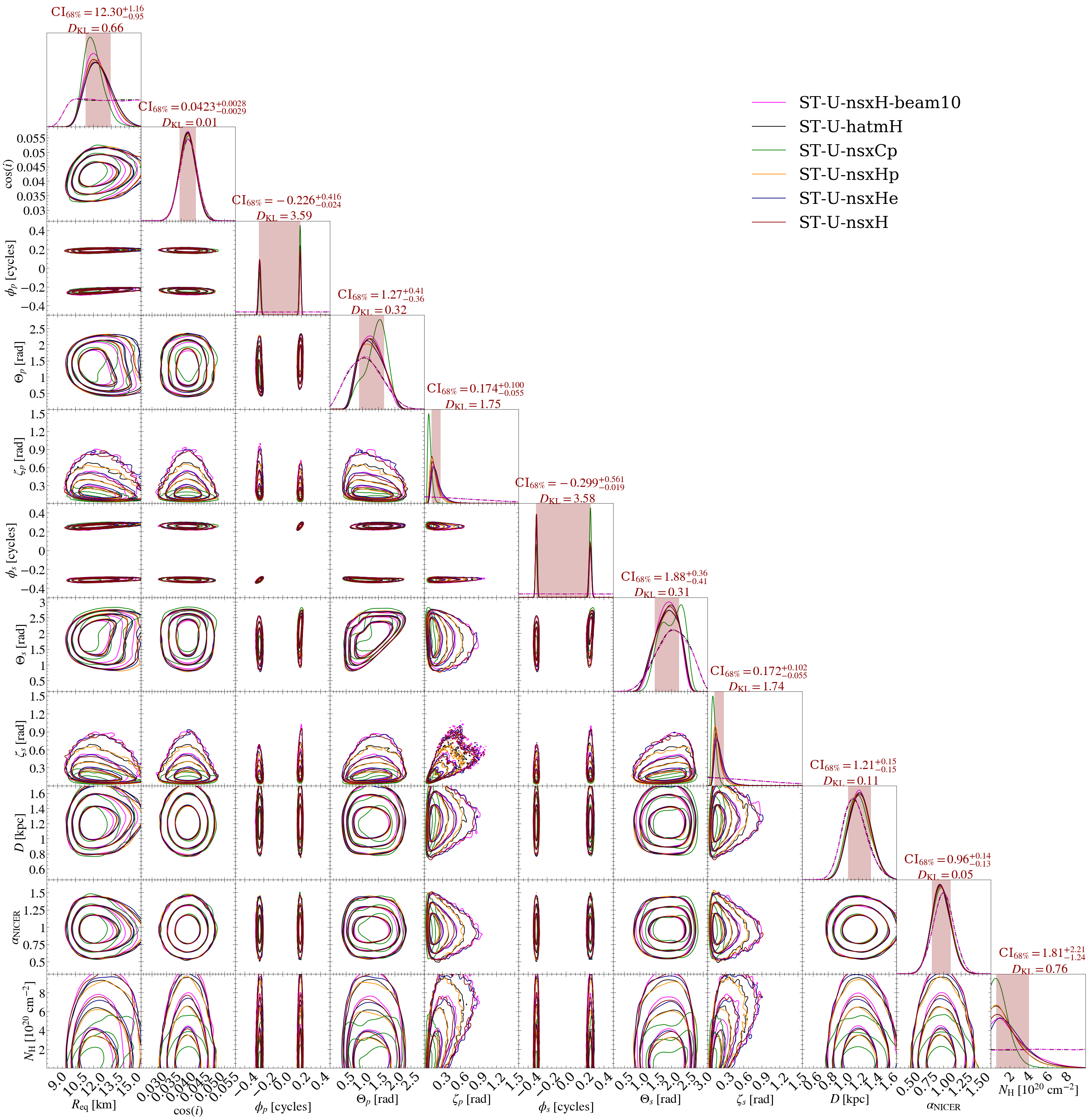}}
\caption{\small{
Effect of the atmosphere model on the posterior distributions of radius and the other parameters not including those shown in Figure \ref{fig:posterior_spacetime_num_J0740} (except the beaming parameters $a$ and $b$) using the \joh \NICER data set conditional on the \texttt{ST-U} model. 
The additional parameters are cosine Earth inclination to spin axis ($\cos i$), primary region initial phase ($\phi_{p}$), primary region center colatitude ($\Theta_{p}$), primary region angular radius ($\zeta_{p}$), secondary region initial phase ($\phi_{s}$), secondary region center colatitude ($\Theta_{s}$), secondary region angular radius ($\zeta_{s}$), distance ($D$), effective area scaling factor of \NICER ($\alpha_{\mathrm{NICER}}$), and interstellar neutral hydrogen column density ($N_{\mathrm{H}}$). 
See the caption of Figure \ref{fig:posterior_spacetime_num_J0740} for more details. %and the Zenodo repository \citet{salmi_zenodo23} for posteriors of the beaming parameters $a$ and $b$.  
}}
\label{fig:posterior_num_J0740_appendix}
\end{figure*}
}

{
\begin{figure*}[t!]
\centering
\resizebox{\hsize}{!}{\includegraphics[
width=\textwidth]{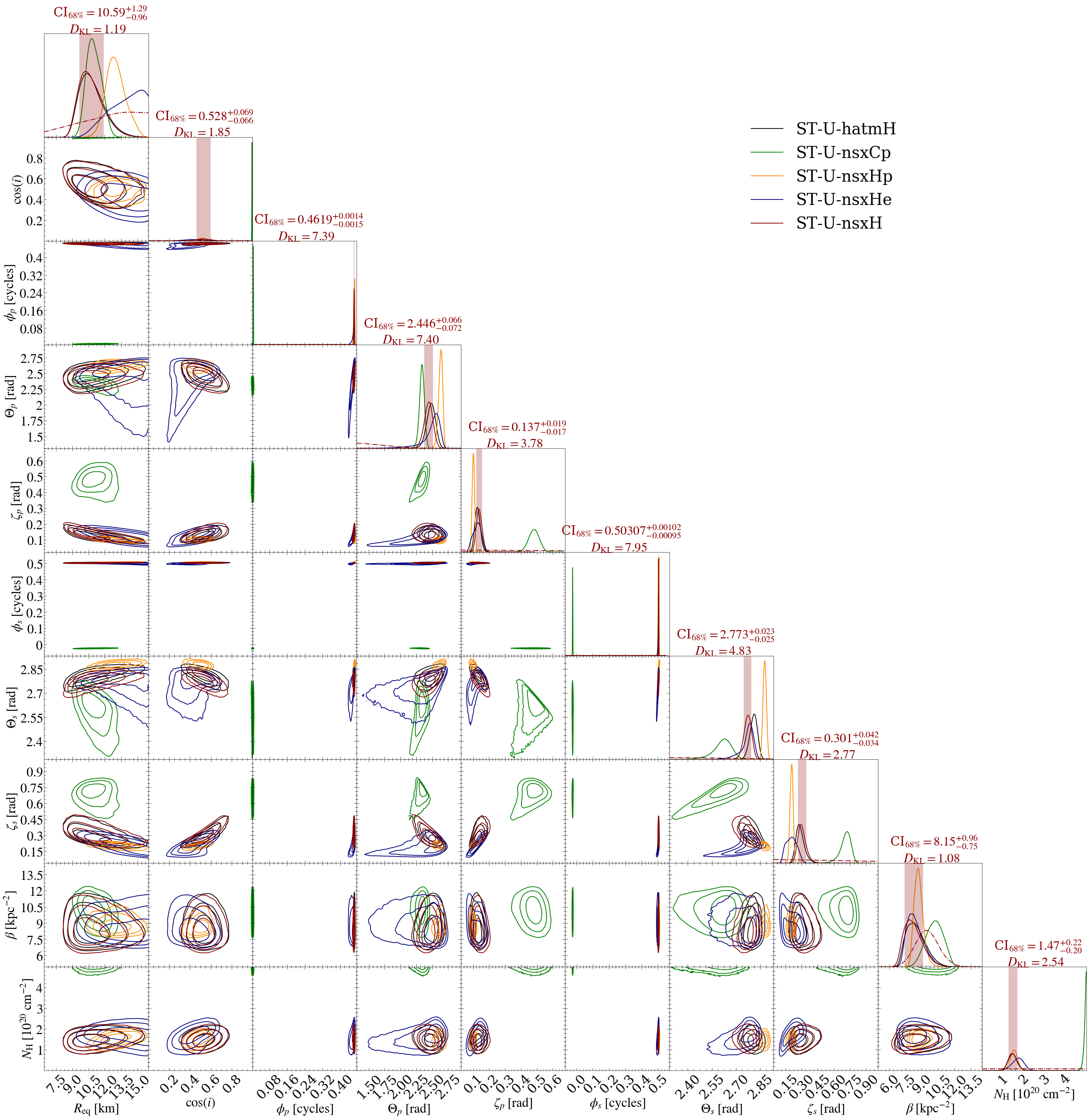}}
\caption{\small{
Effect of the atmosphere model on the posterior distributions of radius and the other parameters not including those shown in Figure \ref{fig:posterior_spacetime_num_J0030} using the \jdbl \NICER data set conditional on the \texttt{ST-U} model. 
The parameters are the same as described in the caption of Figure \ref{fig:posterior_num_J0740_appendix}, except that $D$ and $\alpha_{\mathrm{NICER}}$ are replaced with $\beta = \alpha_{\mathrm{NICER}} D^{-2}$. 
See the caption of Figure \ref{fig:posterior_spacetime_num_J0030} for additional details.
}}
\label{fig:posterior_num_J0030_appendix}
\end{figure*}
}

{
\begin{figure*}[t!]
\centering
\resizebox{\hsize}{!}{\includegraphics[
width=\textwidth]{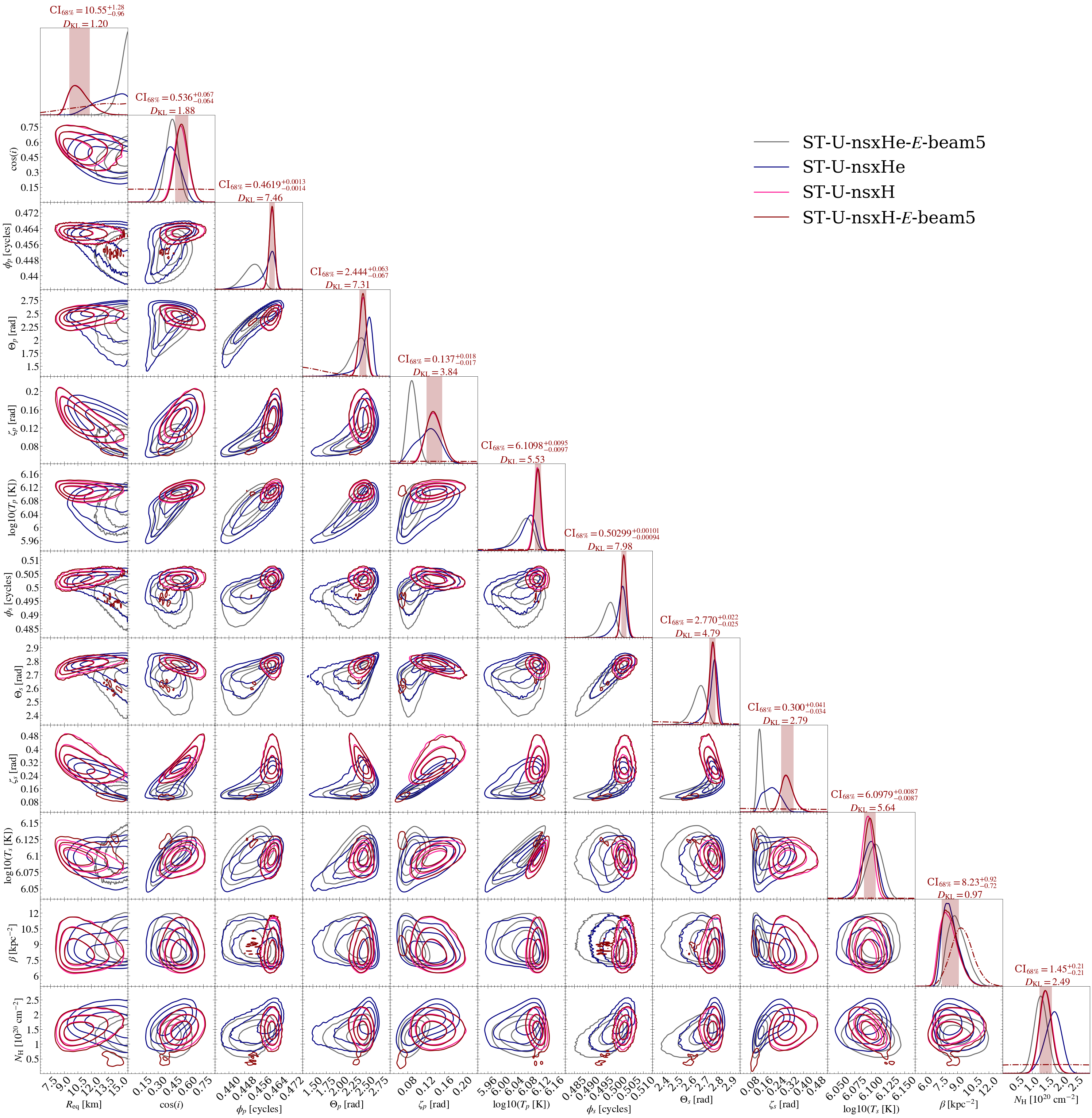}}
\caption{\small{
Effect of the atmosphere model on the posterior distributions of radius and the other parameters not including those shown in Figure \ref{fig:posterior_spacetime_J0030_beam} using the \jdbl \NICER data set conditional on the \texttt{ST-U} model and energy-dependent beaming parameterization. 
The shown parameters are the same as in Figure \ref{fig:posterior_num_J0030_appendix}. 
See the caption of Figure \ref{fig:posterior_spacetime_J0030_beam} for additional details.
}}
\label{fig:posterior_beam_J0030_appendix}
\end{figure*}
}

{
\begin{figure*}[t!]
\centering
\resizebox{\hsize}{!}{\includegraphics[
width=\textwidth]{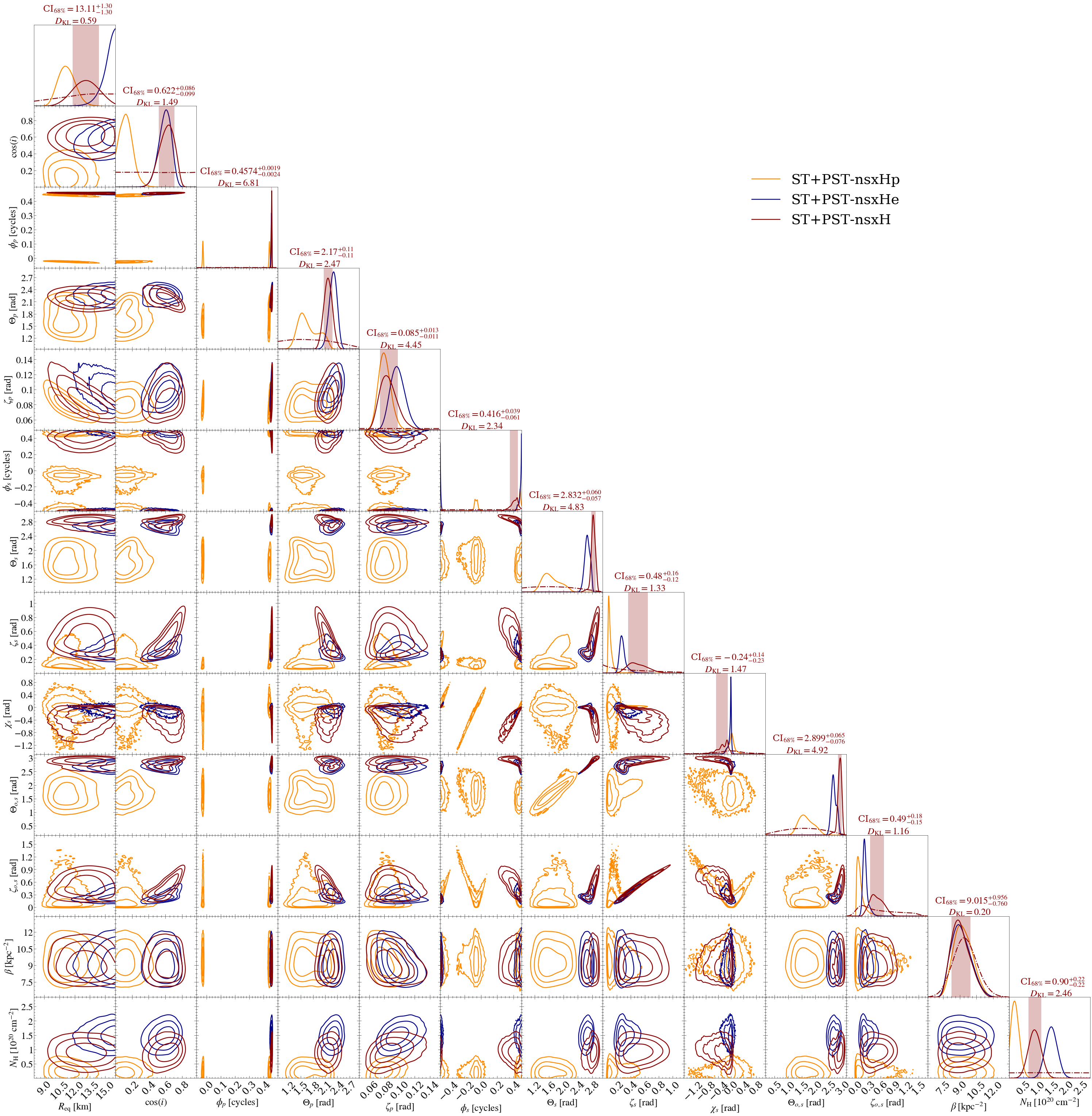}}
\caption{\small{
Effect of the atmosphere model on the posterior distributions of radius and the other parameters not including those shown in Figure \ref{fig:posterior_spacetime_num_J0030_stpst} using the \jdbl \NICER data set conditional on the \texttt{ST+PST} model. 
In addition to the parameters defined in Figures \ref{fig:posterior_num_J0740_appendix} and \ref{fig:posterior_num_J0030_appendix}, we show the posteriors for azimuth offset between the emitting and the masking spherical caps of the secondary hot spot ($\chi_{s}$), colatitude of the masking region of the secondary hot spot ($\Theta_{o,s}$), and radius of the masking region of the secondary hot spot ($\zeta_{o,s}$). 
See the caption of Figure \ref{fig:posterior_spacetime_num_J0030_stpst} for additional details.
}}
\label{fig:posterior_num_J0030_stpst_appendix}
\end{figure*}
}

{
\begin{figure*}[t!]
\centering
\resizebox{\hsize}{!}{\includegraphics[
width=\textwidth]{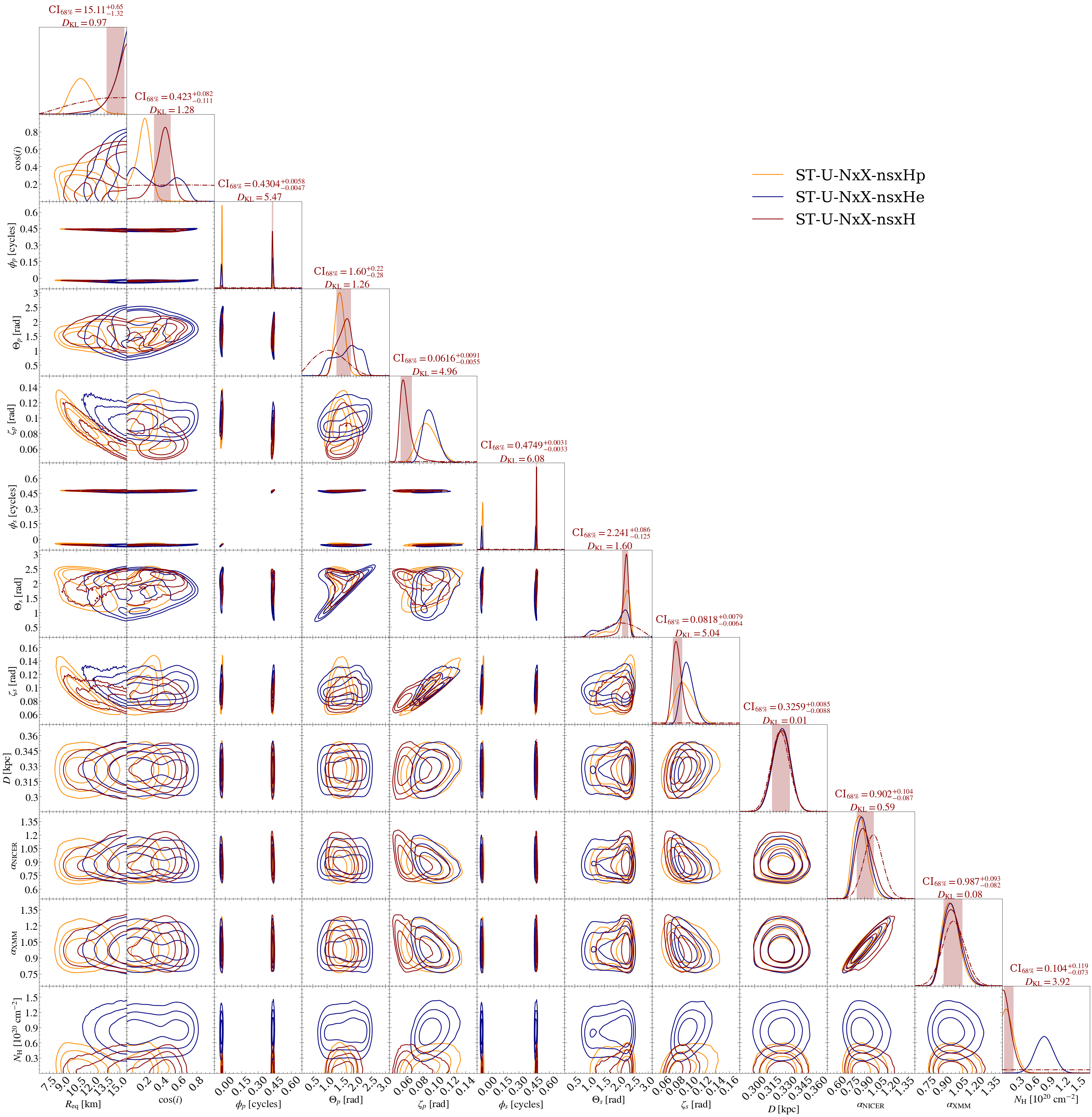}}
\caption{\small{
Effect of the atmosphere model on the posterior distributions of radius and the other parameters not including those shown in Figure \ref{fig:posterior_spacetime_num_J0030_NxX} using the \jdbl joint \NICER and \xmm (NxX) data sets conditional on the \texttt{ST-U} model. 
In addition to the parameters defined in Figure \ref{fig:posterior_num_J0740_appendix}, we show the posteriors for the effective area scaling factor of \xmm ($\alpha_{\mathrm{XMM}}$). 
See the caption of Figure \ref{fig:posterior_spacetime_num_J0030_NxX} for additional details. 
}}
\label{fig:posterior_num_J0030_NxX_appendix}
\end{figure*}
}

\end{document}